\documentclass[aps,onecolumn,tightenlines,amsmath,amssymb,11pt,superscriptaddress,nofootinbib]{revtex4}
\usepackage{graphicx}
\usepackage{amsmath,amssymb,amsfonts,amsthm,stmaryrd,mathtools,mathbbol,bm,physics,tensor}
\usepackage{soul}
\usepackage{xcolor}
\usepackage{tikz}
\allowdisplaybreaks[1]
\usepackage[bookmarks,linktocpage, colorlinks=true, plainpages = false, citecolor = treegreen,  linkcolor=darkblue, urlcolor = darkblue, filecolor = blue]{hyperref} 

\definecolor{tealgreen}{rgb}{0.0, 0.5, 1.0}
\definecolor{darkblue}{rgb}{0., 0.4, 0.8}
\definecolor{cadmiumred}{rgb}{1., 0., 0.22}
\definecolor{treegreen}{rgb}{0., 0.7, 0.3}

\usepackage{pifont}

\def\be#1\ee{\begin{align}#1\end{align}}

\def\ba{\begin{eqnarray}}
\def\ea{\end{eqnarray}}
\def\nn{\nonumber}

\usepackage[normalem]{ulem}

\begin{document}

\title{On  mass inflation and late-time divergences in \\
	limiting-curvature regular black holes
}

\author{Johanna Borissova}
\email{j.borissova@imperial.ac.uk}
\affiliation{Abdus Salam Centre for Theoretical Physics, Imperial College London, London SW7 2AZ, United Kingdom}

\begin{abstract}
\bigskip
{\sc Abstract:} Mass inflation in non-extremal regular black holes with a non-linear dependence of the Misner-Sharp mass $M$ on the asymptotic mass $m>0$ may be a transient phenomenon followed by a model-dependent non-exponential late-time divergence. We analyse these regimes based on the generalised Ori model  for a class of static spherically symmetric and asymptotically flat regular black holes with a de Sitter core and an $m$-independent upper bound on the Misner-Sharp mass density and Kretschmann scalar inspired by quasi-topological gravity. We 
discuss the geometric mechanism behind divergences of  physical quantities
at the inner horizon of  the perturbed spacetime, and show that these require the unphysical mass parameter inside the shell to evolve outside the positive-real domain, or in some cases force a distinct type of dynamical continuation of the static metric.
Imposing an asymptotic late-time divergence in form of a fractional power-law scaling, tunable to be arbitrarily mild within the considered approximation and effectively taming mass inflation, allows us to single out a family of generalised Fan-Wang regular black holes.

\end{abstract}

\maketitle

\tableofcontents

\section{Introduction and setup}\label{Sec:Introduction}

Mass inflation~\cite{Poisson:1989zz,Poisson:1990eh} is a potentially serious obstruction to the physical viability of non-extremal regular black holes with Cauchy horizons~\cite{Brown:2011tv,Carballo-Rubio:2018pmi,Bonanno:2020fgp,Carballo-Rubio:2021bpr,Bonanno:2022rvo,Carballo-Rubio:2022pzu,Bonanno:2023qhp,Bonanno:2025bgc,Carballo-Rubio:2025fnc}. It is typically manifested in an exponential growth of the Misner-Sharp mass $M$ at the inner apparent horizon upon perturbation of the spacetime by ingoing and outgoing fluxes. This results generically in a  local curvature singularity as this inner horizon approaches the Cauchy horizon at asymptotically late times. For regular black holes where $M$ depends non-linearly on the asymptotic mass $m >0$, this conclusion is, however, non-trivial. Within spherical symmetry and under certain assumptions on the radial profile of $M$, the asymptotic late-time behavior  may exhibit a model-dependent non-exponential divergence  following a transient period of mass inflation~\cite{Bonanno:2020fgp,Carballo-Rubio:2021bpr,Bonanno:2022rvo,Carballo-Rubio:2022pzu,Bonanno:2023qhp,Bonanno:2025bgc}.\\

Our aim here is to provide a complementary perspective on mass inflation and late-time divergences for a class of static spherically symmetric and asymptotically flat regular black holes with a de Sitter core and an $m$-independent upper bound on the Misner-Sharp mass density.
 Concretely, we consider 
 regular black holes with Misner-Sharp mass $M$ depending on the asymptotic mass $m > 0$, in the form
 \ba\label{eq:Metricf}
 \dd{s}^2 &=& -f(m,r)\dd{v}^2 + 2 \dd{v}\dd{r}+ r^2 \dd{\Omega}^2\,,
 \ea
 where~\footnote{We work in units where $G_{\text{N}}= c=1$.}
 \ba
f(m,r) &=&  1 - \frac{2 M(m,r)}{r}\,, \quad \quad m \,\,=\,\, \lim_{r\to \infty}M(m,r)\,.
 \ea
We require that the following scalar function, which is one of the invariants characterising the Riemann tensor of the metrics~\eqref{eq:Metricf} and proportional to the Misner-Sharp mass density,
\ba\label{eq:Psi}
\psi(m,r) &\vcentcolon =& \frac{1 -f(m,r)}{r^2} \,\,=\,\, \frac{2 M(m,r)}{r^3}\,,
\ea
as a function of $r \in [0,\infty)$ be universally bounded by inverse powers of a fundamental length scale $\ell$, i.e.,
\ba\label{eq:LimitingCurvaturePsi}
\psi(m,r) & \leq & \psi_{\text{max}} \,\, =\,\, \frac{1}{\ell^2}\,.
\ea
Moreover we assume that this bound is saturated at the core, 
  \ba\label{eq:Psimax}
  \psi_{\text{max}} &=& \lim_{r\to 0} \psi(m,r) \,.
  \ea
  This condition is compatible with the assumption that the regular black holes have a de Sitter core with $\ell$ acting as  an effective regularisation length parameter in an expansion of the metric around $r=0$,
   \ba\label{eq:deSittercore}
 f(m,r) &=&  1 - \psi_{\text{max}}r^2 + \dots \,.
   \ea
  We will later consider specifically examples for which the function $\psi(m,r)$ is strictly monotonically decreasing for $r \in (0,\infty)$ with $\psi'(m,0) = 0$ for $m>0$, where a prime denotes the derivative with respect to the radial coordinate. Hence the limiting Misner-Sharp mass density can be located by the necessary condition
  \ba
  \psi(m,r) & \nearrow & \psi_{\text{max}} \quad \Rightarrow \quad  \psi'(m,r) \,\, \nearrow\,\, 0\,.
  \ea
 Importantly, these assumptions are made for the static background~\eqref{eq:Metricf}. Our main goal is to discuss geometric aspects of mass inflation and related divergences manifested in an unbounded growth of the Misner-Sharp mass density at the inner horizon of the spacetime when perturbed by ingoing and outgoing fluxes, and modeled based on Vaidya generalisations $m\mapsto m(v)$ of the metric~\eqref{eq:Metricf}.\\

Let us emphasise that the assumption of boundedness of the function $\psi(m,r)$ is only a necessary condition for the spacetimes~\eqref{eq:Metricf} to have bounded curvature. In particular, the Kretschmann scalar
\ba
 K  \vcentcolon =  R_{\mu\nu\rho\sigma}R^{\mu\nu\rho\sigma} 
 \ea
for these metrics is given by
\ba\label{eq:Kmr}
K(m,r) &=& 4 \psi(m,r)^2 + \qty[4 \psi(m,r) + 2 r \psi'(m,r)]^2 + \qty[2 \psi(m,r) + 4 r \psi'(m,r) + r^2 \psi''(m,r)]^2\,,\quad 
\ea
and is bounded as a function of $r \in [0,\infty)$ provided all
\ba\label{eq:Scalars}
\qty{\psi(m,r),\, r \psi'(m,r), \, r^2 \psi''(m,r)}
\ea
are bounded on this domain. Throughout we take the notion of regular black hole to refer to boundedness of polynomial curvature invariants, which for static metrics of the form~\eqref{eq:Metricf} will be the case if all scalar quantities~\eqref{eq:Scalars} are bounded. Hence we assume that this is the case for $r\in[0,\infty)$ in the static background spacetime, whereby regularity at $r=0$ is already implied by the de Sitter core~\eqref{eq:deSittercore}. However,  this in general does not mean that a given polynomial curvature invariant such as $K(m,r)$ is bounded by an $m$-independent value, i.e., that the background spacetimes satisfy the limiting-curvature condition
\ba\label{eq:LimitingCurvature}
K(m,r) \,\, \leq \,\, K_{\text{max}} \,\,=\,\, \frac{\alpha}{\ell^4}\,,
 \ea
 where $\alpha$ is a positive numerical coefficient as in Markov's original limiting-curvature conjecture for cosmology~\cite{Markov:1982rcm}. Our examples later will be selected from a particular class of static regular black holes with metric of the form~\eqref{eq:Metricf} which satisfy~\eqref{eq:LimitingCurvaturePsi}, and more generally~\eqref{eq:LimitingCurvature}, and we will come back to this point in Sec.~\ref{Sec:MassInflation}. \\
 
The remainder of this article is structured as follows. In Sec.~\ref{Sec:Ori} we review the generalised Ori model, in which the outgoing flux is modeled by a null thin shell, and collect the main dynamical equations used later to address mass inflation and late-time divergences in regular black holes. In Sec.~\ref{Sec:MassInflation} we focus on a particular class of limiting-curvature regular black holes inspired by quasi-topological gravity. To that end, we first review previous results on mass inflation and late-time scalings for the Hayward regular black hole in Sec.~\ref{SecSub:Hayward}, and subsequently extend these results to a broader class of limiting-curvature regular black holes described by a characteristic algebraic equation in Sec.~\ref{SecSub:QTGClass}. In Sec.~\ref{SecSubSub:FanWang} we discuss a selected example of generalised Fan-Wang regular black holes, whereas in Sec.~\ref{SecSubSub:Other} we show an example featuring a distinct type of mechanism. We finish with a conclusion in Sec.~\ref{Sec:Discussion}.
 
 \section{Review of the generalised Ori model and dynamical equations}\label{Sec:Ori}
 
We will ground the discussion of mass inflation and late-time divergences on a generalisation of the Ori model~\cite{Ori:1991zz} as  in~\cite{Bonanno:2020fgp,Carballo-Rubio:2021bpr,Bonanno:2022rvo,Carballo-Rubio:2022pzu,Bonanno:2023qhp,Bonanno:2025bgc}. To that end, we will first review the main dynamical equations derived in these references.\\

Throughout we assume that the static background regular black hole is non-extremal with two separated horizons. In the Ori model the outgoing perturbation is modeled by a pressureless null thin shell $\Sigma$ placed between the apparent inner and outer horizon. This shell divides the spacetime into an inside region $(+)$ 
an outside region $(-)$.
The metrics in both sectors are assumed to be Vaidya generalisations of~\eqref{eq:Metricf} obtained by promoting the respective mass parameters $m_\pm$ to time-dependent functions, i.e., 
\ba\label{eq:Metricfpm}
\dd{s_\pm^2} &=& - f_\pm\qty(m\qty(v_\pm),r)\dd{v^2_\pm}+ 2 \dd{v_\pm}\dd{r} + r^2 \dd{\Omega}^2\,,
\ea
where
\ba
f(m_\pm\qty(v_\pm),r) &=&  1 - \frac{2 M_\pm\qty(m_\pm\qty(v_\pm),r)}{r}\,,
\ea
and  $M_\pm$ denotes the Misner-Sharp mass in each sector. The radial coordinate $r$ is the same in both sectors by the matching  condition of the induced metrics on $\Sigma$. We will denote the radial position of the shell by $r=R\qty(v_\pm)$. The relation between the two time coordinates $v_\pm$ follows from
\ba\label{eq:vpvmRelation}
f_+\qty(m_+(v_+), R(v_+))\dd{v_+} &=&f_-\qty(m_-(v_-), R(v_-)) \dd{v}_-\,.
\ea
This relation can be used to express all quantities in terms of $v \vcentcolon = v_-$. \\

The continuous ingoing perturbation at asymptotically late times $v\to \infty$ is assumed to follow Price's law~\cite{Price:1971fb,Price:1972pw}, i.e., to decay as a power law
\ba\label{eq:Price}
m_-(v) &\simeq & m_0 - \frac{\beta}{\qty[\frac{v}{v_0}]^{\gamma}}\,, \quad \quad  \gamma \,\,=\,\, p-1 \,\,\geq \,\,11\,,
\ea
where  $p = 12$ corresponds to quadrupole waves and
$m_0 = \lim_{v\to \infty}m_-(v)$ is the asymptotic mass of the static background black hole. Moreover $\beta >0$ is a quantity of dimension mass, or equivalently length in the adopted units, and $v_0$ an initial reference time. The shell approaches the Cauchy horizon at $r=r_-$ of the background regular black hole~\eqref{eq:Metricf} as $v \to \infty$. This is described by the condition
\ba\label{eq:rMinus}
\lim_{v\to \infty}R(v) &=& r_-\,, \quad \quad   \lim_{v\to \infty}f_-\qty(m_-(v),R(v)) \,\, =\,\, f_-\qty(m_0,r_-)\,\,=\,\, 0\,.
\ea
The surface gravity of the Cauchy horizon is given by
\ba\label{eq:kappaMinus}
\kappa_- &=&
 \frac{1}{2}  f_-'\qty(m_0,r_-) \,\, < \,\, 0\,,
\ea
where a prime denotes the derivative with respect to $r$.\\

The independent Einstein  equations in each sector,
\ba
\pdv{M(m(v),r)}{r} \,\,=\,\, - 4 \pi r^2 T\indices{_v^v}\,, \quad \quad \pdv{M(m(v),r)}{v} \,\,=\,\, 4 \pi r^2 T\indices{_v^r}\,,
\ea
as well as continuity of the flux across $\Sigma$, i.e.,~\footnote{Here square brackets indicate the discontinuity of a scalar quantity across the position of the shell.} 
\ba\label{eq:TssJump}
\qty[T_{\mu\nu}s^\mu s^\nu] &=& 0\,, \quad \quad  s_\pm^\mu \,\,=\,\, \qty(\frac{2}{f_\pm},1,0,0)\,,
\ea 
where $s_\pm^\mu$ are the outgoing null normals to $\Sigma$, in combination with~\eqref{eq:vpvmRelation} can be used to infer
\ba
\eval{\frac{1}{f_+\qty(m_+(v),R(v))} \pdv{M_+ \qty(m_+\qty(v),r)}{v}}_{r=R(v)}  &=&  \eval{\frac{1}{f_-\qty(m_-(v),R(v))}\pdv{M_-\qty(m_-\qty(v),r)}{v}}_{r=R(v)} \,.\label{eq:EqM}
\ea
This relation can be further rewritten as
\ba\label{eq:EqM2}
 \frac{1}{f_+\qty(m_+(v),R(v))} \pdv{M_+\qty(m_+(v),R(v))}{m_+(v)} \dot{m}_+(v) &=&   \frac{1}{f_-\qty(m_-(v),R(v))} \pdv{M_-\qty(m_-(v),R(v))}{m_-(v)} \dot{m}_-(v) \,,\quad 
\ea
where a dot stands for the derivative with respect to $v$. The equation of motion for the radial position of the shell can be obtained from the radial null condition 
$f_- \dd{v}_- =2 \dd{r}$
evaluated at $r=R(v)$. \\

In summary, introducing the definitions~\footnote{The definition of the function $F\qty(m_-(v),R(v))$ is equivalent to the one introduced in~\cite{Bonanno:2020fgp,Bonanno:2022rvo,Bonanno:2025bgc}, whereas the second function is equal to $G(m_+(v),R(v))= g(m_+(v),R(v)) f_+\qty(m_+(v),R(v))$,  where the function denoted by $g(m_+(v),R(v)) $ in~\cite{Bonanno:2025bgc} is defined as the first factor in~\eqref{eq:G}.}
\ba
F(m_-(v),R(v)) &\vcentcolon = &  \frac{1}{f_-\qty(m_-(v),R(v))} \pdv{M_-\qty(m_-(v),R(v))}{m_-(v)} \dot{m}_-(v)\,,\label{eq:F}\\
G(m_+(v),R(v))  & \vcentcolon=& \qty[\pdv{M_+\qty(m_+(v),R(v))}{m_+(v)}]^{-1} f_+\qty(m_+(v),R(v)) \,,\label{eq:G}
\ea
the dynamics of the generalised Ori model is encoded in the two equations
\ba
\dot{R}(v) \,\,=\,\,\frac{1}{2}f_-\qty(m_-(v),R(v))\,, \quad \quad  \dot{m}_+(v) \,\,=\,\, G(m_+(v),R(v)) F(m_-(v),R(v))  \label{eq:mrdot}\,,
\ea
which represent a coupled set of differential equations determining the radial position of the shell $R(v)$ and the mass parameter $m_+(v)$ once $m_-(v)$ has been prescribed as in~\eqref{eq:Price}. Therefrom the Misner-Sharp mass and the Kretschmann scalar evaluated at the position of the shell can be obtained,
\ba\label{eq:Mv}
M_+\qty(v) \,\, \vcentcolon = \,\, M_+\qty(m_+(v),R(v))\,, \quad \quad  K_+(v) \,\, \vcentcolon = \,\, K\qty(m_+(v),R(v))\,.
\ea
The above equations hold for a generic background~\eqref{eq:Metricf}.~\footnote{Notice that these equations are obtained by realising the static background regular black hole as an effective non-vacuum solution of general relativity. These same static regular black holes can be realised as vacuum solutions of modified gravitational theories, so-called four-dimensional general quasi-topological gravities~\cite{Borissova:2026krh,Borissova:2026wmn,Bueno:2025zaj,Colleaux:2017ibe,Colleaux:2019ckh,Colleaux:2026hat}.
	Mass inflation based on thin-shell models in related higher-dimensional polynomial curvature quasi-topological gravities~\cite{Oliva:2010eb,Myers:2010jv,Dehghani:2011vu,Cisterna:2017umf,Bueno:2019ltp,Bueno:2019ycr, Bueno:2022res,Moreno:2023rfl,Bueno:2024dgm,Frolov:2024hhe,Bueno:2024zsx,Bueno:2024eig,Bueno:2025gjg,Bueno:2025qjk} has been addressed e.g.~in~\cite{Frolov:2026rcm,DiFilippo:2026jpv,Bueno:2026wjt}.
} \\

The asymptotic late-time dynamics of the Misner-Sharp mass can be analysed based on an approximate solution for $R(v)$ in the limit $v\to \infty$~\cite{Bonanno:2020fgp,Carballo-Rubio:2021bpr,Bonanno:2025bgc}. To that end, $m_-(v)$ and $R(v)$ are expanded around their background values,
\ba
m_-(v) \,\,=\,\, m_0 + \delta m_-(v)\,, \quad \quad  R(v) \,\,=\,\, r_- + \delta R(v)\,,
\ea
and in turn the metric function $f_-\qty(m_-(v),R(v))$ is expanded around these values. The first equation of motion in~\eqref{eq:mrdot} together with~\eqref{eq:rMinus} and~\eqref{eq:kappaMinus} imply
\ba
\derivative{\delta R(v) }{v}&=& 
 \frac{1}{2} \pdv{f_-\qty(m_0,r_-)}{m_0} \delta m_-(v) + \frac{1}{2}  \pdv{f_-\qty(m_0, r_-)}{r_-} \delta R(v) +  \mathcal{O}\qty(\delta m_-^2, \delta m_- \delta R, \delta R^2)\nn\\
  &=\vcentcolon & - \frac{1}{r_-} A(m_0,r_-)  \delta m_-(v) +  \kappa_- \delta R(v) + \mathcal{O}\qty(\delta m_-^2, \delta m_- \delta R, \delta R^2)\,,\label{eq:DeltaRdot}
  \ea
 where
 \ba
  A(m_0,r_-) \,\,\vcentcolon =\,\, \pdv{M_-(m_0,r_-)}{m_0}
\ea
is a dimensionless model-dependent parameter.
Neglecting the subleading terms in~\eqref{eq:DeltaRdot} and inserting $\delta m_-(v) = - \frac{\beta}{(v/v_0)^\gamma}$ according to Price's law~\eqref{eq:Price}, the solution of~\eqref{eq:DeltaRdot} is given by 
\ba
\delta R(v) & \simeq &  e^{\kappa_- v} \qty[c - \frac{A \beta}{r_- \kappa_-} v_0^\gamma \kappa_-^\gamma \Gamma[1-\gamma, \kappa_- v]]
\,\,\simeq \,\,  -  \frac{A \beta}{r_- \kappa_-} \qty[\frac{v}{v_0}]^{-\gamma} \qty[1 - \frac{\gamma}{ \kappa_- v} 
 +\mathcal{O}\qty(\frac{1}{v^2})]\,,\label{eq:DeltaRSol}
\ea
where $c$ is an integration constant which, importantly, does not enter the expansion for $v\to \infty$.~\footnote{The final expression in~\eqref{eq:DeltaRSol} is equivalent to~$(12)$ in {\it v2} of~\cite{Carballo-Rubio:2021bpr} and~$(67)$ in {\it v1} of~\cite{Bonanno:2025bgc} taking into account the definition of the Frobenius coefficient in~$(37a)$ in this latter reference and adapted to our conventions $a_0 \vcentcolon =  \frac{A \beta}{r_- \kappa_-}$.}
Using moreover
\ba
\dot{m}_-(v) &=& \frac{\beta \gamma}{v_0} \qty[\frac{v}{v_0}]^{-(\gamma+1)}\,,
\ea
it follows that to leading orders
\ba\label{eq:rdotSol}
\dot{R}(v) &=&  \frac{A}{r_- \kappa_- }  \dot{m}_-(v) \qty[1- \frac{\gamma +1}{ \kappa_- v} +\mathcal{O}\qty(\frac{1}{v^2})] \,\,=\,\,  \frac{\lambda}{v^{\gamma+1}} \qty[1- \frac{\gamma +1}{ \kappa_- v} +\mathcal{O}\qty(\frac{1}{v^2})] \,,
\ea
where $\lambda$ is a constant parameter with length dimension $\gamma+1$ defined by
\ba\label{eq:Lambda}
 \lambda \qty(m_0,r_-,\kappa_-,\beta,\gamma,v_0) &  \vcentcolon = & \frac{A\qty(m_0,r_-)}{r_- \kappa_-} \beta \gamma v_0^\gamma\,.
\ea
Inserting this result into the definition of the function $F\qty(m_-(v),R(v))$ in~\eqref{eq:F}, with $f_-\qty(m_-(v),R(v))$ expressed in terms of $\dot{R}(v)$ using the first equation of motion~\eqref{eq:mrdot}, the derivatives of $m_-(v)$ cancel out, leading to~\footnote{This relation is equivalent to~$(30)$ in {\it v1} of~\cite{Carballo-Rubio:2021bpr} and~$(69)$ in {\it v1} of~\cite{Bonanno:2025bgc} taking into account the different notations and conventions of these references.}
\ba\label{eq:FSol}
F\qty(m_-(v),R(v)) &=& \frac{1}{2\dot{R}(v)}  \pdv{M_-\qty(m_-(v),R(v))}{m_-(v)} \dot{m}_-(v)
\,\, \simeq \,\, \qty[\frac{2 A}{r_- \kappa_-} \qty[1- \frac{\gamma+1}{\kappa_- v} + \mathcal{O}\qty(\frac{1}{v^2})]]^{-1} A\quad \quad \nn\\
&\simeq &  \frac{r_-}{2}\qty[ \kappa_- + \frac{\gamma+1}{ v} + \mathcal{O}\qty(\frac{1}{v^2})]\,.
\ea
 In particular this function approaches a constant non-zero value as $v\to \infty$,
\ba
 \lim_{v \to \infty} F\qty(m_-(v),R(v)) &=& \frac{r_- \kappa_-}{2}\,.
\ea
The previous identities hold independently of the background geometry~\eqref{eq:Metricf}. 
\\

The above result can be used, instead of the equation of motion for $m_+(v)$, to derive an approximate equation describing the evolution of the Misner-Sharp mass $M_+(v)$ at late times~\cite{Carballo-Rubio:2021bpr}. To that end, defining the dimensionless quantity
\ba\label{eq:B}
B\qty(m_+\qty(v),R(v)) & \vcentcolon = & \pdv{M_+ \qty(m_+\qty(v),R(v))}{R(v)} \,, 
\ea
notice first that
\ba
\dot{M}_+(v) &=&  \pdv{M_+ \qty(m_+\qty(v),R(v))}{m_+(v)} \dot{m}_+(v)  +  \pdv{M_+ \qty(m_+\qty(v),R(v))}{R(v)} \dot{R}(v)\nn\\
&=&f_+\qty(m_+(v),R(v)) F\qty(m_-(v), R(v)) + B\qty(m_+\qty(v),R(v))\dot{R}(v)\,,\label{eq:MdotExact}
\ea
where~\eqref{eq:EqM2} and the definition~\eqref{eq:F} are used. This relation is exact, in particular instead of~\eqref{eq:mrdot} one may equivalently solve the system
\ba
\dot{R}(v) &=&  \frac{1}{2}f_-\qty(m_-(v),R(v))\,,\label{eq:Mrdot1}\\
 \dot{M}_+(v) &=&  f_+\qty(M_+(v),R(v)) F\qty(m_-(v), R(v)) + B\qty(M_+(v),R(v))\dot{R}(v) \,,\label{eq:Mrdot2}
\ea
where the function defined in~\eqref{eq:B} must be written  in terms of $M_+(v)$ and $R(v)$.~\footnote{Depending on the metric, this requires in general an explicit algebraic inversion of $M_+\qty(m_+(v),R(v))$ in the form $m_+\qty(M_+(v),R(v))$
which may have several solution branches or be ill-defined at isolated points. In this case, the resulting master equation~\eqref{eq:MdotExact} and its approximation~\eqref{eq:Mdot} can be applied within a given domain where the function $M_+\qty(m_+(v),R(v))$ is locally invertible.
}
 Using~\eqref{eq:Metricfpm},~\eqref{eq:rdotSol} and~\eqref{eq:FSol}, the last equation can be approximated as
\ba
\dot{M}_+(v) &\simeq &  -\qty[ \kappa_- + \frac{\gamma+1}{ v} + \mathcal{O}\qty(\frac{1}{v^2})]\qty[ M_+(v)- \frac{r_-}{2} ] +\frac{ \lambda}{v^{\gamma+1}} B\qty(M_+(v),R(v))\qty[1+ \mathcal{O}\qty(\frac{1}{v})] \,.\quad \label{eq:Mdot}
\ea
\\

The first term linear in $M_+(v)$ in~\eqref{eq:Mdot} triggers an initially exponential growth (in magnitude) of the solution, i.e., standard mass inflation,
 while the second term may alter this behavior at asymptotically late times depending on the quantity $B\qty(M_+(v),R(v))$. The case when this function is proportional to positive integer powers
$  M_+\qty(v)^{n+1}
$ for $n \geq 1 $,
with a possibly $R(v)$-dependent prefactor which is approximated as $R(v) \simeq r_-$ at late times, has been discussed in~\cite{Carballo-Rubio:2021bpr} and leads to a polynomial divergence of $M_+(v)$ for $v\to \infty$. In the next section we will discuss concrete examples of this type which satisfy the condition of limiting Misner-Sharp mass density~\eqref{eq:LimitingCurvaturePsi} for the static background.~\footnote{Notice that the equation for $M_+(v)$ in~\eqref{eq:Mrdot2} and its approximation~\eqref{eq:Mdot} hold for a generic choice of exterior and interior metrics $f_\pm\qty(m_\pm(v),r)$. The choice of the exterior metric enters only through the function $F\qty(m_-(v),R(v))$ which appears in the first term and governs the exponential mass-inflation regime. A possible departure from standard mass inflation at late times is determined entirely by the choice of the interior metric through the quantity $B\qty(M_+(v),R(v))$ defined in~\eqref{eq:B}.}

 \section{Mass inflation and late-time divergences in limiting-curvature regular black holes
}\label{Sec:MassInflation}

\subsection{Review: Hayward regular black hole}\label{SecSub:Hayward}

 We will now first reproduce previous results on mass inflation and the late-time polynomial scaling of the Misner-Sharp mass in the context of the generalised Ori model applied to the Hayward regular black hole~\cite{Bonanno:2020fgp,Carballo-Rubio:2021bpr,Bonanno:2022rvo,Carballo-Rubio:2022pzu,Bonanno:2023qhp,Bonanno:2025bgc}. The Hayward metric~\cite{Hayward:2005gi} takes the form~\eqref{eq:Metricf} with
\ba\label{eq:fHayward}
f(m,r) &=& 1 - \frac{2 m r^2 }{r^3 + 2 m \ell^2}\,.
\ea
This  function in general has three roots and describes a non-extremal regular black hole with two horizons on the domain $r \in [0,\infty)$  provided $m > m_{\text{crit}} =\frac{3 \sqrt{3}}{4}\ell $. We will denote the outer and inner horizon of this black hole by $r=r_\pm $. The remaining root of $f(m,r)$ in this case is located at a negative $r=r_0 $  determined by $r_- r_+ r_0= - 2 m \ell^2$. \\

For  positive mass parameter $m > 0$ and $r \in (0,\infty)$, the Misner-Sharp mass density~\eqref{eq:Psi} of the Hayward spacetime~\eqref{eq:fHayward},
\ba\label{eq:PsiHayward}
 \psi(m,r) &=& \frac{2 m}{r^3 + 2 m \ell^2}\,,
\ea 
is a strictly monotonically decreasing function and satisfies the limiting condition~\eqref{eq:LimitingCurvaturePsi} 
with a universal $m$-independent bound given by
\ba\label{eq:LimitingCurvatureHaywardPsi}
\psi_{\text{max}} &=& \lim_{r\to 0} \psi(m,r) \,\,=\,\, \frac{1}{\ell^2}\,.
\ea
Under these assumptions, one may also verify from the
expression for the Kretschmann scalar,
\ba\label{eq:KHayward}
K(m,r)&=& \frac{4 8 m^2}{\qty(r^3 + 2 m\ell^2 )^6} \qty[r^{12} - 8 m \ell^2 r^9
+ 72  m^2 \ell^4r^6 - 16m^3 \ell^6  r^3 + 32m^4 \ell^8 ]\,,
\ea
that the limiting-curvature condition~\eqref{eq:LimitingCurvature} is satisfied with a universal $m$-independent upper bound given by
\ba\label{eq:LimitingCurvatureHayward}
K_{\text{max}} &=& \lim_{r\to 0}K(m,r)\,\,=\,\, \frac{24}{\ell^4}\,.
\ea 
The same conclusions apply to the dynamical Hayward spacetime obtained by replacing $m \mapsto m(v)$ in~\eqref{eq:fHayward}, on the domain $r\in [0,\infty)$ and assuming positivity of the mass function $m(v)$. In fact the expressions~\eqref{eq:PsiHayward} and~\eqref{eq:KHayward} for the Vaidya-generalised Hayward spacetime take the same form with $m=m(v)$, and do not  depend on time derivatives.\\

 For negative pass parameter $m < 0$ and $r \in [0,\infty)$, on the other hand, the Hayward metric~\eqref{eq:fHayward} does not describe a regular black hole and, importantly, exhibits a singularity in the Misner-Sharp mass density and Kretschmann scalar at a finite $ r=r_* $ given by~\footnote{Mathematically, the Hayward metric is
	invariant under the CPT-type transformation
	\ba\label{eq:Transformation}
	v \,\,\to\,\,  -v\,, \quad \quad r \,\, \to \,\, - r\,,\quad \quad  m \,\, \to \,\,-m\,,
	\ea
	while the Misner-Sharp mass transforms as $M(-m,-r) = -M(m,r)$. The condition $f(r,m)=f(-r,-m)$ has been discussed as a consistency constraint for regular black holes in four-dimensional gravitational theories with second-order equations of motion on spherically symmetric backgrounds~\cite{Borissova:2026wmn,Borissova:2026krh,Bueno:2025zaj,Colleaux:2017ibe,Colleaux:2019ckh,Colleaux:2026hat}, which is necessary for  the dynamical evolution of matter across a curvature-regular core~\cite{Thaalba:2026abz}.}
\ba\label{eq:HaywardSingularity}
r_*^3 &=& - 2 m \ell^2 \,\,=\,\, 2 \abs{m} \ell^2  \quad   \Rightarrow  \quad   \lim_{r \to r_*} \psi(-\abs{m},r) \,\, \to \,\, \pm \infty\,,
\ea
where the sign of the divergence depends on the direction from which it is approached. 
This is the only divergence of $\psi(m,r)$ and $K(m,r)$ as functions of $ r \in [0,\infty)$  which a Hayward metric can exhibit. 
On these grounds, one may anticipate that mass inflation triggering an unbounded growth of the Misner-Sharp mass $M_+(v)$ defined in~\eqref{eq:Mv} requires the mass parameter $m_+(v)$ to become negative dynamically.
Moreover, the fact that
  $R(v) \simeq r_-$ for $v \to \infty$ implies that any, not necessarily exponential, asymptotic late-time divergence of the Misner-Sharp mass requires the variable $m_+(v)$ to approach a negative finite value singled out by the physical singularity~\eqref{eq:HaywardSingularity} of the negative-mass Hayward spacetime, i.e.,
  \ba\label{eq:mStarHayward}
 m_{+*}  & \vcentcolon =&  \lim_{v\to \infty}m_+(v) \,\,= \,\, - \frac{r_-^3}{2 \ell^2}\,\, < \,\, 0\,.
  \ea
  That the late-time dynamics derived from the generalised Ori model is indeed governed by such a quasi-fixed point of the system $\qty(R(v),m_+(v))$ has been discussed in~\cite{Bonanno:2020fgp,Carballo-Rubio:2021bpr,Bonanno:2022rvo,Carballo-Rubio:2022pzu,Bonanno:2023qhp,Bonanno:2025bgc}, and is reconfirmed numerically and analytically below. In the next subsection we will extend these observations to a broader class of regular black holes with limiting Misner-Sharp mass density and curvature.\\

\begin{figure}[h!]
	\centering
	\includegraphics[width=0.65\textwidth]{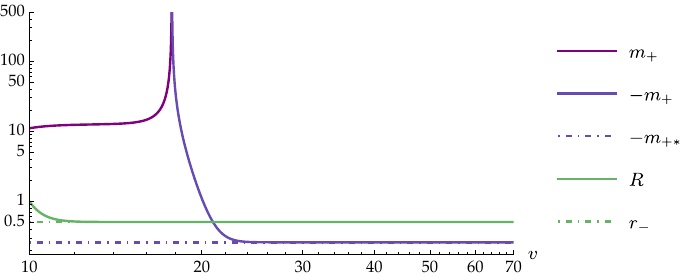}
	\caption{\label{Fig:HaywardOri} Radial position of the shell $R(v)$ and mass parameter $m_+(v)$ obtained by numerically integrating~\eqref{eq:mdotSolHayward} with free parameters set to $m_0 = 10$, $\ell = \frac{1}{2}$, $\beta = 1$, $\gamma=11$ and $v_0 =10$, and initial values $R(v_0)=1$ and $m_+(v_0)=10+1$, in appropriate units of an external reference length scale. The dashed lines mark the finite constant values $R(v)\simeq r_-$ and $m_+(v) \simeq m_{+*}$ given by~\eqref{eq:mStarHayward}, which are approached asymptotically as $v\to \infty$.}
\end{figure}

\begin{figure}[h!]
	\centering
	\includegraphics[width=0.65\textwidth]{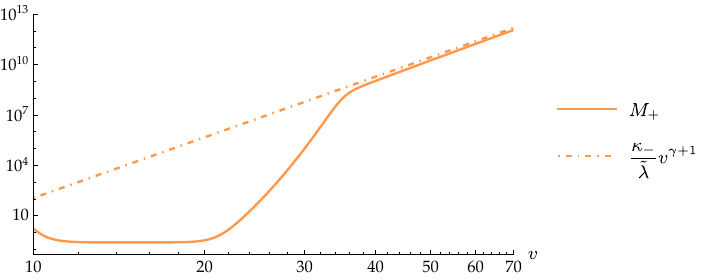}
	\caption{\label{Fig:HaywardOri2} Misner-Sharp mass $M_+(v)$ for the configuration corresponding to the choice of free parameters and initial values specified in Fig.~\ref{Fig:HaywardOri}, and analytically determined polynomial late-time scaling~\eqref{eq:MHaywardSol} following the transient mass-inflation period.}
\end{figure}

\begin{figure}[h!]
	\centering
	\includegraphics[width=0.65\textwidth]{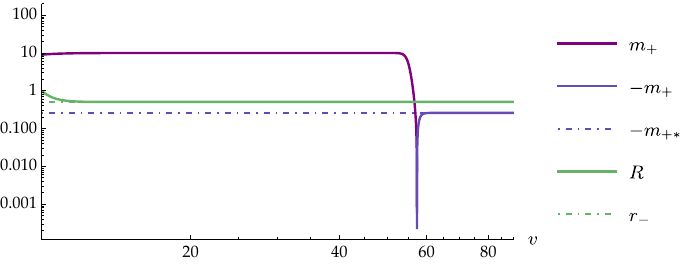}
	\caption{\label{Fig:HaywardOriSecond} Radial position of the shell $R(v)$ and mass parameter $m_+(v)$ obtained by numerically integrating~\eqref{eq:mdotSolHayward} with free parameters set to $m_0 = 10$, $\ell = \frac{1}{2}$, $\beta = 1$, $\gamma=11$ and $v_0 =10$, and initial values $R(v_0)=1$ and $m_+(v_0)=10-1$, in appropriate units of an external reference length scale. 
	}
\end{figure}

\begin{figure}[h!]
	\centering
	\includegraphics[width=0.65\textwidth]{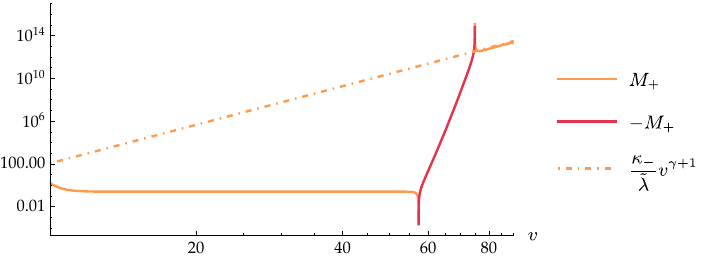}
	\caption{\label{Fig:HaywardOriSecond2} Misner-Sharp mass $M_+(v)$ for the configuration corresponding to the choice of free parameters and initial values specified in Fig.~\ref{Fig:HaywardOriSecond}, and analytically determined polynomial late-time scaling~\eqref{eq:MHaywardSol}.}
\end{figure}

We verify the previous expectations by applying the derivation in Sec.~\ref{Sec:Ori} with Misner-Sharp mass functions given by
\ba\label{eq:fHaywardpm}
 M_\pm \qty(m_\pm(v),R(v)) &=&  \frac{ m_\pm\qty(v) R(v)^3 }{R(v)^3 + 2 m_\pm\qty(v) \ell^2}\,.
\ea
The equations of motion~\eqref{eq:mrdot} are explicitly
\ba
\dot{R}(v)\,\,=\,\, \frac{1}{2} - \frac{m_-(v)R(v)^2}{R(v)^3 + 2 m_-(v)\ell^2} \,, \quad \quad 
 \dot{m}_+(v) \,\,=\,\, G\qty(m_+(v),R(v))  F\qty(m_-(v),R(v))\,,
\label{eq:mrdotHayward}
\ea
where
\ba
G\qty(m_+(v),R(v)) &=&  R(v)^{-6}\qty[R(v)^3 + 2 m_+(v)\ell^2 ]\qty[R(v)^3 - 2 m_+(v) \qty(R(v)^2-\ell^2)]\,,\\
F\qty(m_-(v),R(v))&=& \dot{m}_-(v) R(v)^6 \qty[R(v)^3 + 2 m_-(v)\ell^2 ]^{-1}\qty[R(v)^3 - 2 m_-(v) \qty(R(v)^2-\ell^2)]^{-1}\,,
\ea
with $m_-(v)$ given by~\eqref{eq:Price}.
Fig.~\ref{Fig:HaywardOri} and Fig.~\ref{Fig:HaywardOriSecond} show the result of the numerical integration of~\eqref{eq:mrdotHayward} for $R(v)$ and $m_+(v)$ for a fixed choice of free parameters and two different types of initial conditions distinguished by $m_+\qty(v_0) = m_0 \pm 1$. 
In the former case the solution for $m_+(v)$ ranges in two disjoint intervals $\qty(m_+(v_0),\infty)$ and $\qty(-\infty,m_{+*})$, whereas in the latter case it ranges in a single interval $\qty(m_{+*}, m_+(v_0))$, where $m_{+*}<0$ is defined  in~\eqref{eq:mStarHayward}.~\footnote{In practice, since for the first scenario shown in Fig.~\ref{Fig:HaywardOri}, $m_+(v)$ exhibits a discontinuity at a finite $v$, the numerical integration is performed in terms of the variable $s_+(v) \vcentcolon= 1/m_+(v)$. Alternatively, one may directly integrate~\eqref{eq:Mrdot1} and~\eqref{eq:Mrdot2} for this type of initial conditions and use the inverse of  $M_+\qty(m_+(v),R(v))$ in the form $m_+(v) = m_+\qty(M_+(v),R(v))$.} 
From the solutions for $R(v)$ and $m_+(v)$, the Misner-Sharp mass $M_+(v)$ can be computed according to~\eqref{eq:fHaywardpm} and is shown in Fig.~\ref{Fig:HaywardOri2} and Fig.~\ref{Fig:HaywardOriSecond2}, respectively, where we also compare the late-time regime with the analytically derived power-law expression below. Qualitatively similar plots were produced originally in~\cite{Bonanno:2020fgp,Carballo-Rubio:2021bpr,Bonanno:2022rvo,Carballo-Rubio:2022pzu,Bonanno:2023qhp,Bonanno:2025bgc}. \\

For the first type of initial conditions considered in Fig.~\ref{Fig:HaywardOri} and Fig.~\ref{Fig:HaywardOri2}, $M_+(v)$ stays positive and finite at all finite times. By contrast, the second type of initial conditions considered in Fig.~\ref{Fig:HaywardOriSecond} and Fig.~\ref{Fig:HaywardOriSecond2} triggers $M_+(v)$ to cross zero, to stay negative during mass inflation, and to run into a discontinuity at a finite $v$ as the system transitions into the universal late-time  positive power-law regime which is independent of the initial conditions, as is reviewed below. While the question of initial conditions is in general external input outside of the regime of validity of the Ori model, in the remainder of this article we take the viewpoint that configurations exhibiting a divergence of the physical Misner-Sharp mass and curvature within a finite time are physically not viable in addressing the question of eternal late-time divergences associated with a Cauchy horizon. In later sections will only consider initial conditions for which $M_+(v)$ evolves continuously up to $v\to \infty$. These will be selected by the requirement that the  surface energy density of the outgoing null shell between the two horizons, computed from the future-directed null generators $n_\pm^\mu = -s_\pm^\mu$~\cite{Barrabes:1991ng},
\ba\label{eq:Mu}
\mu &=& \frac{\qty[M]}{4 \pi R^2}\,,
\ea
is positive, i.e., by the condition $\qty[M_+- M_-](v_0) > 0$. An analogue perspective has been advocated in~\cite{Carballo-Rubio:2021bpr,Carballo-Rubio:2022pzu}, cf.~\cite{Bonanno:2020fgp,Carballo-Rubio:2021bpr,Bonanno:2022rvo,Carballo-Rubio:2022pzu,Bonanno:2023qhp,Bonanno:2025bgc} for a debate on this and related questions. 
 \\

\begin{figure}[h!]
	\centering
	\includegraphics[width=0.65\textwidth]{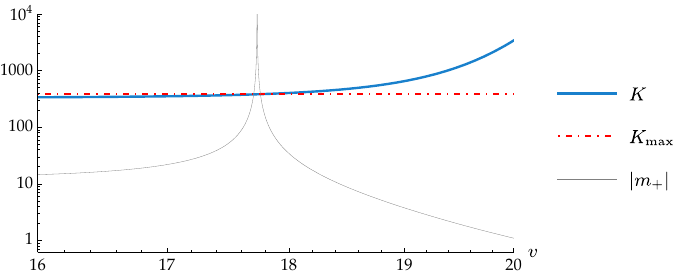}
	\caption{\label{Fig:HaywardOri3} Kretschmann scalar $K_+(v)$ for the configuration corresponding to the choice of free parameters and initial values specified in Fig.~\ref{Fig:HaywardOri}, zoomed into the regime of onset of mass inflation. As $K_+(v)$ grows past $K_{\text{max}}$ of the static background, the mass parameter $m_+(v)$ encounters a discontinuity and becomes instantaneously negative dynamically.}
\end{figure}

\begin{figure}[h!]
	\centering
	\includegraphics[width=0.65\textwidth]{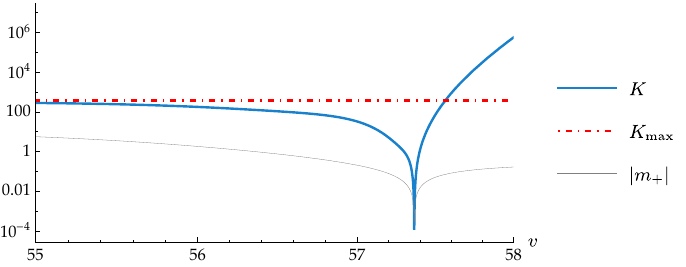}
	\caption{\label{Fig:HaywardOriSecond3} Kretschmann scalar $K_+(v)$ for the configuration corresponding to the choice of free parameters and initial values specified in Fig.~\ref{Fig:HaywardOriSecond}, zoomed into the regime of onset of mass inflation. As the mass parameter $m_+(v)$ changes sign and becomes sufficiently negative dynamically, $K_+(v)$ is unhaltered to grow beyond $K_{\text{max}}$ of the static background. 
	}
\end{figure}

The asymptotic late-time dynamics of the Misner-Sharp mass $M_+(v)$ can be elegantly discussed based on the approximated master equation~\eqref{eq:Mdot}~\cite{Carballo-Rubio:2021bpr}. For the Hayward metric~\eqref{eq:fHayward}, the definitions in~\eqref{eq:B} and~\eqref{eq:Lambda} result in
\ba
B\qty(M_+(v),R(v)) &=&  \frac{6 \ell^2}{R(v)^4}M_+(v)^2 \,\, \simeq \,\,\frac{6 \ell^2}{r_-^4}M_+(v)^2\,,\\
\lambda &=&  \frac{r_-^5 \beta \gamma v_0^\gamma}{\qty(r_-^3 + 2 m_0 \ell^2)^2 \kappa_-} \,\, \simeq \,\, \frac{r_- \beta \gamma v_0^\gamma}{4 m_0^2 \kappa_-}\,\,< 0\,,
\ea
where the last approximation holds for $\ell \ll m_0$, in which case $r_- \simeq \ell$.
Then the differential equation~\eqref{eq:Mdot}, neglecting subleading terms in the limit $v\to \infty$ and the constant term in the second square brackets of the first term, becomes
\ba\label{eq:MdotHayward}
\dot{M}_+\qty(v) & \simeq & - \kappa_-M_+(v) + \frac{\tilde{\lambda}}{v^{\gamma+1}} M_+(v)^2\,,
\ea
where $\tilde{\lambda} \vcentcolon= \frac{6 \ell^2}{r_-^4}\lambda <0$ is a constant parameter with length dimension $\gamma-1$. Assuming $M_+(v)>0$, this is a Bernoulli differential equation with solution given by~\footnote{For the first type of initial conditions considered in Fig.~\ref{Fig:HaywardOri} and Fig.~\ref{Fig:HaywardOri2}, $M_+(v)$ stays positive throughout the evolution and this analytic derivation of the late-time scaling~\eqref{eq:MHaywardSol} can be directly applied. For the second type of initial conditions considered in Fig.~\ref{Fig:HaywardOriSecond} and Fig.~\ref{Fig:HaywardOriSecond2}, however, $M_+(v)$ becomes and stays negative during mass inflation. This can be used from~\eqref{eq:MdotHayward}, together with the fact that $\tilde{\lambda} <0$, to show that the solution encounters a divergence at a finite $v$, as is indeed the case numerically from Fig.~\ref{Fig:HaywardOriSecond2}. The derivation of the universal positive-divergent  late-time scaling~\eqref{eq:MHaywardSol} from~\eqref{eq:MdotHayward} is then applicable on a disjoint domain for $M_+(v)$ following this discontinuity, if such a dynamical evolution can be made sense of physically.}
\ba\label{eq:MHaywardSol}
M_+(v) & \simeq & \frac{e^{- \kappa_- v} }{c + \tilde{\lambda} \kappa_-^\gamma \Gamma\qty[-\gamma, \kappa_- v]} \,\, \simeq \,\, \frac{\kappa_-}{\tilde{\lambda}}v^{\gamma+1} + \mathcal{O}\qty(v^\gamma)\,,
\ea
where $c$ is an integration constant which, notably, does not enter the asymptotic late-time expansion. This result can be used, moreover, to obtain the late-time asymptotics of the mass parameter $m_+(v)$ using~\eqref{eq:fHaywardpm}, which gives 
\ba\label{eq:mExpansionHayward}
m_+(v) &=& m_{+*} - \frac{r_-^6 }{4  \ell^4 } \frac{\tilde{\lambda}}{\kappa_-}\frac{1}{v^{\gamma+1}} + \mathcal{O}\qty(\frac{1}{v^{2 \qty(\gamma+1)}})\,.\label{eq:mdotSolHayward}
\ea 
This approach of directly solving the differential equation for $M_+(v)$ automatically selects the attractive quasi-fixed point~\eqref{eq:mStarHayward} out of a possible pair of quasi-fixed points of the dynamical system $\qty(R(v),m_+(v))$ discussed in~\cite{Bonanno:2020fgp,Bonanno:2022rvo,Bonanno:2025bgc}.~\footnote{For completeness, we note that the Frobenius coefficient denoted by $a_0$ in~\cite{Bonanno:2025bgc} and identified as $a_0 \vcentcolon= \frac{A\beta}{r_- \kappa_-}$ in our conventions has been approximated in these references for the Hayward case as
\ba
a_0 &=& \frac{r_-^5 \beta}{\qty(r_-^3 + 2 \ell^2 m_0)^2\kappa_-}  \,\, \simeq \,\, \frac{r_- \beta}{4 m_0^2 \kappa_-}\,,
\ea
assuming $\ell\ll m_0$, in which case $r_- \simeq \ell$, cf.~$(68b)$ in {\it v1}  of~\cite{Bonanno:2025bgc} or~$(23)$ in {\it v3} of~\cite{Bonanno:2020fgp}.} 
From~\eqref{eq:mdotSolHayward} one may also compute the Kretschmann scalar~\eqref{eq:Mv} to see that it diverges as $K_+(v) \sim v^{6(\gamma+1)}$ in the limit $v\to \infty$. Instead of  showing the Kretschmann scalar on the full domain of numerical integration, which looks qualitatively similar to behaviour of the Misner-Sharp mass in Fig.~\ref{Fig:HaywardOri2} and Fig.~\ref{Fig:HaywardOriSecond2}, in Fig.~\ref{Fig:HaywardOri3} and Fig.~\ref{Fig:HaywardOriSecond3} we show a plot zooming into the regime of onset of mass inflation, to illustrate that the growth of $K_+(v)$ past $K_{\text{max}}$ is released once the mass parameter $m_+(v)$ becomes sufficiently negative.\\

In summary, we have revisited mass inflation and the late-time regime for the Hayward regular black hole~\cite{Bonanno:2020fgp,Carballo-Rubio:2021bpr,Bonanno:2022rvo,Carballo-Rubio:2022pzu,Bonanno:2023qhp,Bonanno:2025bgc}  based on the purely geometric expectation that $m_+(v)$ becomes negative at late times, and from this perspective confirmed that this unphysical mass variable approaches a finite negative value corresponding to the physical singularity of the negative-mass Hayward spacetime~\eqref{eq:HaywardSingularity}. Both of these phenomena are necessary in order accommodate the growth (in magnitude) of  the physical Misner-Sharp mass density and curvature past their maximal values of the positive-mass unperturbed background, and in order to ensure their eternal late-time divergence as $v\to \infty$. In the next subsection we will extend these observations to a broader class of limiting-curvature regular black holes.

\subsection{Quasi-topological class of limiting-curvature regular black holes}\label{SecSub:QTGClass}

The Hayward regular black hole is a particular example of a more general class of  asymptotically flat double-horizon de Sitter core regular black holes with metric~\eqref{eq:Metricf} satisfying the condition of limiting Misner-Sharp mass density~\eqref{eq:LimitingCurvaturePsi}, and more generally limiting curvature~\eqref{eq:LimitingCurvature}, for which the former as a function of $r\in (0,\infty)$ can be written as
\ba\label{eq:PsiHRho}
\psi(m,r) &=& H\qty(\varrho(m,r)) \,, \quad \quad \varrho(r,m) \,\, = \,\, \frac{2m}{r^3}  \quad  \Rightarrow  \quad
M(m,r) \,\,=\,\, \frac{r^3}{2} H\qty(\varrho(m,r))\,,
\ea
initially with the domains  for the background spacetime given by
\ba\label{eq:DomainsStatic}
\psi(m,r) &\in & \qty(0,\psi_{\text{max}})\,, \quad \quad \varrho(r,m) \,\,\in\,\, (0,\infty)\,,
\ea
and the limit $r\to 0$ prescribed in~\eqref{eq:Psimax}, i.e., $\lim_{r\to 0} \psi(m,r)=\psi_{\text{max}}$.
Schwarzschild asymptotics for $r\to \infty$ requires $H(\varrho) \simeq \varrho$ in the limit $\varrho \to 0$.  The scalar quantities~\eqref{eq:Scalars} for such spacetimes can be expressed as
\ba
r \psi'(m,r) &=&  - 3 \varrho H'(\varrho)\,, \quad \quad 
r^2 \psi''(m,r) \,\,=\,\,  3 \varrho \qty[4  H'\qty(\varrho) + 3  \varrho H''(\varrho)]\,,
\ea
so that boundedness of the Kretschmann scalar $K(m,r)$  in~\eqref{eq:Kmr} for $r\in (0,\infty)$ translates into boundedness of
\ba
\qty{H(\varrho), \, \varrho H'(\varrho),\, \varrho^2 H''(\varrho)}
\ea
on the domain $\varrho \in (0,\infty)$.  To ensure boundedness for $r \to 0$, we require that $H(\varrho)$ be bounded as $\varrho \to \infty$, while its derivative $H'(\varrho)$ is assumed to decay faster than $\sim\frac{1}{\varrho}$ and its second derivative $H''(\varrho)$ is assumed to decay faster than $\sim \frac{1}{\varrho^2}$ for $\varrho \to \infty$. For related considerations see~\cite{Bueno:2026dln,Hennigar:2025yqm}. 
The Kretschmann scalar of such a de Sitter core regular black hole,
\ba 
K(m,r) &=& 4 H(\varrho)^2 +  \qty[4 H(\varrho)  - 6 \varrho H'(\varrho)]^2 + \qty[2H(\varrho)  + 9  \varrho^2 H''(\varrho)]^2\,,\quad 
\ea
then does not depend on $m$, and hence has an $m$-independent upper bound as observed in~\cite{Frolov:2024hhe,Bueno:2026dln}. To see this explicitly, notice that $H$ has the same dimension as $\varrho$ and $\psi_{\text{max}} =\frac{1}{\ell^2}$, which is length dimension $-2$. Hence one may define dimensionless variables and functions~\footnote{For simplicitly we assume that the metric does not depend on additional length scales beyond the ones introduced so far.}
\ba
\hat{\varrho} &=& \ell^2 \varrho\,,\quad \quad \derivative{}{\varrho} \,\,=\,\, \frac{1}{\ell^2}\derivative{}{\hat{\varrho}}\,, \quad \quad \hat{H} \,\,=\,\, \ell^2 H\,,
\ea
to write
\ba 
K(m,r) &=& \frac{1}{\ell^4}\qty[4 \hat{H}(\hat{\varrho})^2 +  \qty[4\hat{H}(\hat{\varrho})  - 6 \hat{\varrho} H'(\hat{\varrho})]^2 + \qty[2\hat{H}(\hat{\varrho})  + 9  \hat{\varrho}^2 \hat{H}''(\hat{\varrho})]^2]\,.\quad 
\ea
The quantity in large square brackets is just a positive dimensionless number for any given $\hat{\varrho} \in (0,\infty)$. Whatever its bound is, the actual scale is set by the fundamental length scale $\ell$ independently of the mass $m$.
In this sense, such Sitter core regular black holes satisfy the limiting-curvature condition~\eqref{eq:LimitingCurvature}, including the condition of limiting Misner-Sharp mass density~\eqref{eq:LimitingCurvaturePsi}.\\

Invoking now explicitly our working assumptions in Sec.~\ref{Sec:Introduction}, the requirement that the limiting value of $\psi(m,r)$ is attained as $r\to 0$ translates into~\footnote{This also implies $\lim_{r\to 0}K(m,r) = 24 \lim_{\varrho \to \infty}H(\varrho)^2 = 24 \psi_{\text{max}}^2$, consistently e.g.~with~\eqref{eq:LimitingCurvatureHayward} for the Hayward spacetime.} 
\ba \label{eq:PsimaxH}
\psi_{\text{max}} &=& \lim_{\varrho \to \infty} H(\varrho)\,.
\ea
The assumption that $\psi(m,r)$ is strictly monotonically decreasing for $r\in(0,\infty)$, i.e., $\psi'(m,r) < 0$, after taking the radial derivative on both sides of the defining relation~\eqref{eq:PsiHRho}, implies
\ba
H'(\varrho) &>& 0\,,
\ea
for $\varrho \in (0,\infty)$. Hence the function $H(\varrho)$ has an inverse on this domain, which will be denoted by $h(\psi)$, i.e.,
\ba\label{eq:HRho}
\psi \,\,=\,\, H(\varrho)   \quad  \Leftrightarrow  \quad h(\psi) &=& \varrho \,,
\ea
for $\psi \in (0,\psi_{\text{max}})$, and moreover on this domain
\ba
h'(\psi) &=& \frac{1}{H'(\varrho)} \,\, > \,\, 0\,.
\ea
The limiting Misner-Sharp mass density~\eqref{eq:PsimaxH}
corresponds to a pole of this characteristic function,
\ba\label{eq:hPsimax}
\lim_{\psi \nearrow \psi_{\text{max}}}h(\psi) &=& \infty \,.
\ea 
Translating the condition on the first derivative of the function $H(\varrho)$ in the limit $\varrho \to \infty$ into a condition on  $h(\psi)$ and its first derivative gives
\ba\label{eq:hPsihPrimemax}
\lim_{\varrho \to \infty} \varrho H'(\varrho) \searrow 0 \,\,\, \quad \,\,\, \Rightarrow \,\,\, \quad \,\,\, \lim_{\psi \nearrow \psi_{\text{max}}}\frac{h(\psi)}{h'(\psi)} & \searrow & 0\,.
\ea
The previous identities provide a useful diagnosis in determining later when the Misner-Sharp mass density of the perturbed spacetime has reached $\psi_{\text{max}}$ of the static background. \\

In summary, below we will consider static asymptotically flat de Sitter core regular black holes with asymptotic mass $m>0$ which can be described by an algebraic equation~\footnote{Such regular black holes can  be obtained as vacuum solutions of four-dimensional non-polynomial curvature quasi-topological gravities~\cite{Bueno:2025zaj,Borissova:2026wmn,Borissova:2026krh,Borissova:2026prd,Borissova:2026klg,Borissova:2026rbi,Konoplya:2026gim,Liu:2026iou}, in analogy to regular black hole vacuum solutions of higher-dimensional polynomial curvature quasi-topological
gravities~\cite{Bueno:2024dgm,Frolov:2024hhe,Bueno:2024zsx,Bueno:2024eig,Bueno:2025gjg,Bueno:2025tli,Bueno:2025qjk, Hennigar:2025yqm,Aguayo:2025xfi,Tsuda:2026xjc,Bueno:2026dln}. In this work, however, we are realising these  spacetimes as effective non-vacuum solutions of general relativity. This is in particular the case when making use of the junction conditions in the discussion of the generalised Ori model in Sec.~\ref{Sec:Ori}. The bulk relation between spherically symmetric vacuum solutions of quasi-topological gravities and generalised Vaidya solutions of general relativity is discussed in~\cite{Borissova:2026prd,Borissova:2026dlz}.}
\ba\label{eq:hmAlgebraic}
h\qty(\frac{1-f}{r^2}) &=& \frac{2 m}{r^3} \,,
\ea
where $h(\psi)$ is an invertible function defined on the domain $\psi \in (0,\psi_{\text{max}})$, has a pole on approach of the upper boundary $\psi \to \psi_{\text{max}}$, with $h'(\psi) > 0$ on this domain and $h'(0)=1$. It is important to keep in mind, however, that the geometry is kinematically defined by prescribing the function $H(\varrho)$ in~\eqref{eq:PsiHRho}, and existence and form of an inverse $h(\psi)$, in the sense of~\eqref{eq:HRho}, on a domain other than $\varrho \in (0,\infty)$ must be addressed separately.
\\

We will now consider perturbing such regular black holes in the framework of the generalised Ori model reviewed in Sec.~\ref{Sec:Ori}. 
Our main  objective is to discuss aspects of the geometric mechanism by which the (magnitude of the) variable proportional to the Misner-Sharp mass density,
\ba\label{eq:PsiDef}
\psi_+(v) &\vcentcolon =& \psi_+(m_+(v),R(v)) \,\, =\,\, \frac{2M_+(v)}{R(v)^3}\,,
\ea
evolves dynamically beyond $\psi_{\text{max}}$, and continues to grow unbounded at asymptotically late times. These two regimes of  surpassing the limiting value of the background and maintaining a divergence for $v\to \infty$ are a priori strictly separate regimes.

When addressing these questions, we will impose on a physically viable solution that  $\psi_+(v)$ be 
differentiable and hence in particular continuous at all times.~\footnote{E.g.,~in the context of the Hayward regular black hole discussed in Sec.~\ref{SecSub:Hayward}, this means we are discarding initial conditions leading to configurations of the type as in Fig.~\ref{Fig:HaywardOriSecond} and Fig.~\ref{Fig:HaywardOriSecond2}, for which $M_+(v)$ grows negative exponentially during mass inflation and encounters a discontinuity before eventually reaching the universal late-time power-law solution.} Our goal is particularly to emphasise conditions on the unphysical mass density variable
\ba\label{eq:Rhop}
\varrho_+(v) &\vcentcolon = & \varrho_+\qty(m_+(v),R(v))\,,
\ea
which are necessary for a divergence of the physical Misner-Sharp mass density $\psi_+(v)$ in the limit $v\to \infty$.  
Essentially, the main point is that by our assumptions on the static limiting-curvature background, the function $ H\qty(\varrho) \in \qty(0,\psi_{\text{max}})$ is finite for $\varrho \in (0,\infty)$. Thus a growth (in magnitude) of $\psi_+(v) = H\qty(\psi_+(v))$ beyond $\psi_{\text{max}}$ can be expected to release an evolution of $\varrho_+(v)$ 
 outside this original domain. Alternatively, such a growth may be associated with a non-trivial dynamical continuation of the spacetime in which the function $H\qty(\psi_+(v))$ enters a new solution branch where it can grow unbounded despite $\varrho_+(v)  \in (0,\infty)$. We will investigate specific examples for both types of phenomena later. \\

To discuss the late-time dynamics of $\psi_+(v)$, we will make use of the approximated master equation for the evolution of the Misner-Sharp mass $M_+(v)$ in~\eqref{eq:Mdot}, as well as the solution for $\dot{R}(v)$ in~\eqref{eq:rdotSol}.
The function defined in~\eqref{eq:B} for the above class of backgrounds is given by 
\ba
B\qty(m_+(v),R(v)) &=& \pdv{M_+(m_+(v),R(v))}{R(v)} \,\,=\,\, \frac{3 M_+(v)}{R(v)} - \frac{3 m_+(v)}{R(v)} H'\qty(\frac{2 m_+(v)}{R(v)^3})\,.\label{eq:Bh}
\ea
The model-dependent background parameter $\lambda$ defined in~\eqref{eq:Lambda} is explicitly
\ba\label{eq:LambdaH}
\lambda &=& \frac{1}{r_- \kappa_-} \pdv{M_-\qty(m_0,r_-)}{m_0} \beta \gamma v_0^\gamma \,\, =\,\, \frac{\beta \gamma v_0^\gamma}{r_- \kappa_-} H'\qty(\frac{2 m_0}{r_-^3})  \,\, < \,\, 0\,,
\ea
where we have used the fact that $H'\qty(\varrho) > 0$ for $\varrho \in (0,\infty)$.  Equation~\eqref{eq:Mdot} and its rewriting below requires in general that $M_+\qty(m_+(v),R(v))$  can be locally inverted for $m_+\qty(M_+(v),R(v))$, or equivalently, for this class of examples, that $H\qty(\varrho_+(v)) = \psi_+(v)$ can be locally inverted in the form $\varrho_+(v) = h\qty(\psi_+(v))$, so that we can write
\ba\label{eq:BPsi}
B\qty(\psi_+(v),R(v)) &=& \frac{3}{2} R(v)^2\qty[\psi_+(v) - \frac{h\qty(\psi_+(v))}{h'\qty(\psi_+(v))}]\,.
\ea
 Keeping this in mind, let us see how far we can get. Using the expression for the Misner-Sharp mass in~\eqref{eq:PsiHRho} and invoking~\eqref{eq:Mdot}, dropping the constant term in the second square brackets, leads to
	\ba
	\dot{M}_+(v) & = & 
	\frac{3}{2} R(v)^2 \dot{R}(v) H\qty(\varrho_+(v)) + \frac{1}{2}R(v)^3 H'\qty(\varrho_+(v)) \dot{\varrho}_+(v)\nn\\
	& \stackrel{!}{\simeq} &-\qty[\kappa_- + \frac{\gamma+1}{v}]M_+(v) + \frac{\lambda}{v^{\gamma+1}}B\qty(m_+(v),R(v)) \nn\\
	&=& - \qty[\kappa_- + \frac{\gamma+1}{v}] \frac{R(v)^3}{2} H\qty(\varrho_+(v)) + \frac{\lambda}{v^{\gamma+1}} \frac{3}{2} R(v)^2  \qty[H\qty(\varrho_+(v)) - \varrho_+(v)H'\qty(\varrho_+(v))] \,.\quad \quad
	\ea
Approximating $\dot{R}(v) \simeq \frac{\lambda}{v^{\gamma+1}}$ according to~\eqref{eq:rdotSol}, one term cancels out on both sides, and we obtain
	\ba\label{eq:Psidot}
	\dot{\psi}_+(v) &=& 
	  H'\qty(\varrho_+(v)) \dot{\varrho}_+(v) 
	\,\, \simeq \,\,   - \qty[\kappa_- + \frac{\gamma+1}{v}]  H\qty(\varrho_+(v)) - \frac{3 \lambda}{r_- v^{\gamma+1}} \qty[ \varrho_+(v)H'\qty(\varrho_+(v))]\,,
	\ea
	where $H\qty(\varrho_+(v)) = \psi_+(v)$. Assuming $H'\qty(\varrho_+(v))$ is non-zero, we may divide the previous equation by this quantity to obtain a differential equation for $\varrho_+(v)$ in the late-time regime,
	\ba\label{eq:Rhodot}
\dot{\varrho}_+(v) &\simeq & 
  - \qty[\kappa_- + \frac{\gamma+1}{v}] \frac{ H\qty(\varrho_+(v))}{H'\qty(\varrho_+(v))} - \frac{3 \lambda}{r_- v^{\gamma+1}}  \varrho_+(v)\,.
\ea	
Alternatively, assuming $\psi_+(v)\neq 0$, we may rewrite~\eqref{eq:Psidot} as
	\ba\label{eq:Psidot2}
	\frac{\dot{\psi}_+(v)}{\psi_+(v)} &\simeq &  - \qty[\kappa_- + \frac{\gamma+1}{v}]   - \frac{3 \lambda}{r_- v^{\gamma+1}} \qty[ \frac{\varrho_+(v)H'\qty(\varrho_+(v))}{H\qty(\varrho_+(v))}]\nn\\
	&=&    - \qty[\kappa_- + \frac{\gamma+1}{v}]- \frac{3 \lambda}{r_- v^{\gamma+1}} \qty[\frac{h\qty(\psi_+(v))}{ \psi_+(v)h'\qty(\psi_+(v))} ]\,,
	\ea
	where the second line applies whenever $H\qty(\varrho_+(v)) = \psi_+(v)$ is locally invertible in a given range of $\psi_+(v) \in \mathbb{R}$, with inverse which we will continue to denote by $h\qty(\psi_+(v))$ as in~\eqref{eq:HRho}. It is important, however, to keep in mind that existence and form of this inverse have to be addressed on possibly disjoint domains separately.
	Neglecting the second term for the moment, the solution produces the standard mass inflation scenario,
	\ba\label{eq:MassInflation}
	\psi_+(v) & \simeq & \frac{ \psi_0 \,e^{- \kappa_-(v-v_0)}}{\qty[\frac{v}{v_0}]^{\gamma+1}}\,,
	\ea
	with some finite initial value $\psi_0 = \psi(v_0)$ at a time when the above approximation applies. The physical quantity $\psi_+(v)$ can in principle be negative or positive during mass inflation,  and this depends on the initial conditions imposed when solving the exact system, as we have seen e.g.~for the Hayward case in Sec.~\ref{SecSub:Hayward}. If $\psi_+(v)$, initially required to be positive, has become negative by crossing zero,
then $\varrho_+(v)$ must have simultaneously touched zero, cf.~e.g.~Fig.~\ref{Fig:HaywardOriSecond} and Fig.~\ref{Fig:HaywardOriSecond2} for an example in the Hayward case. In this case, the second line in~\eqref{eq:Psidot2} is meaningful only with $h\qty(\psi_+(v))$ denoting the inverse of $H\qty(\varrho_+(v))$ on whatever new domain $\varrho_+(v)$ is evolved to.
 On the other hand, if $\psi_+(v)$ remains positive 
and grows  to reach $\psi_+(v_*)=\psi_{\text{max}}$ at some finite time $v=v_*$, then $\varrho_+(v) \to \infty$ for $v \nearrow v_*$ by~\eqref{eq:hPsimax}. This sort of divergence of the unphysical mass variable inside the shell was already observed explicitly for the Hayward spacetime, cf.~e.g.~Fig.~\ref{Fig:HaywardOri} and Fig.~\ref{Fig:HaywardOri3}.

Eventually, however, and in a regime generally (in magnitude) well above $\psi_{\text{max}}$, the second term in~\eqref{eq:Psidot} and~\eqref{eq:Rhodot} may alter the exponential behavior~\eqref{eq:MassInflation}. Notably, in the first representation~\eqref{eq:Psidot2} this term depends only on the ratio between the function $h\qty(\psi_+(v))$ and its derivative, while in the second representation~\eqref{eq:Rhodot} the modification enters linearly in $\varrho_+(v)$. 
We will now formulate a conjecture on this model-dependent function associated with an eternal late-time divergence of 
 $H\qty(\varrho_+(v))$ as $v\to \infty$, and to that end concretely distinguish the following two scenarios.\\

The first scenario is that the second term in~\eqref{eq:Psidot} for large $\psi_+(v)$ contributes by a leading term linear in $\psi_+(v)$, 
	\ba\label{eq:Massinflationeternal}
 \lim_{\psi_+(v)\to \infty}	\frac{h\qty(\psi_+(v))}{h'(\psi_+(v))} \,\,\simeq \,\, \gamma \psi_+(v) + o\qty(\psi_+(v))\,,
	\ea
	where $\gamma$ is a numerical constant.
	 When~\eqref{eq:Massinflationeternal} holds, we expect  the solution to remain exponentially divergent for $v\to \infty$, i.e., mass inflation in such types of limiting-curvature regular black holes would be eternal. This is a model-dependent condition which can be directly checked later based on the generating function for a given example, even on the ground of a naive extension of $h(\psi)$ beyond $(0,\psi_{\text{max}})$. Of course, whether this expectation actually holds must be checked numerically by integrating the full system of equations. 
	\\

The second scenario is that $\psi_+(v)$ grows 
subexponentially for $v\to \infty$, in which case
	\ba
\lim_{v\to \infty} \frac{\dot{\psi}_+(v)}{\psi_+(v)}
 \,\, =\,\, 0\,.
\ea
According to~\eqref{eq:Psidot2}, for this to be the case, the second term must counteract the first one in the large-$\psi_+(v)$ regime. This
entails a universal late-time scaling of the following quantity,
\ba\label{eq:UniversalScaling}
\frac{\varrho_+(v) H'(\varrho_+(v))}{H(\varrho_+(v))} &=&  - \frac{r_- \kappa_-}{3\lambda } v^{\gamma+1}+ \mathcal{O} \qty(v^\gamma) \,\, < \,\, 0\,.
\ea
Notice that the right-hand side is  negative. This identity can therefore only apply somewhere outside the original domain~\eqref{eq:DomainsStatic} characterising the static background. Indeed, remember that $\varrho H'\qty(\varrho) >0$ aside from $H\qty(\varrho)>0$ for $\varrho \in (0,\infty)$. The above relation then yields a necessary  condition for such a subexponential late-time behavior of the solution, which we can write as
\ba\label{eq:SingDetect}
\lim_{\psi_+(v)\to \infty} \frac{h\qty(\psi_+(v)) }{\psi_+(v) h'\qty(\psi_+(v))} &=& -\infty \,.
\ea
It is intuitive to expect that this condition represents the aimed-for criterion of, in effect, reaching a curvature singularity of the static background existing somewhere upon extension beyond $\varrho \in (0,\infty)$. We will investigate this for an example below.
In fact the above condition is intertwined with the assumption that  the dynamics of $\varrho_+(v)$ in the limit $v\to \infty$ is described by a quasi-fixed point~\cite{Bonanno:2025bgc,Bonanno:2022rvo,Bonanno:2020fgp}, i.e.,
\ba\label{eq:RhoFP}
\lim_{v\to \infty} \dot{\varrho}_+(v) &\to & 0\,, \quad \quad \varrho_{+*} \,\,\vcentcolon = \,\, \lim_{v \to \infty}\varrho_+(v) \,.
\ea
Assuming $\varrho_{+*}$ is finite,  using~\eqref{eq:Rhodot} or~\eqref{eq:SingDetect}, this requires
\ba \label{eq:RhoFPConditionH}
\lim_{\varrho_+(v) \to \varrho_{+*}}\frac{H\qty(\varrho_+(v))}{\varrho_+(v) H'\qty(\varrho_+(v))} &=& 0 \quad \Rightarrow \quad \lim_{\psi_+(v) \to \infty}\frac{\psi_+(v) h'\qty(\psi_+(v))}{h\qty(\psi_+(v))} \,\, = \,\, 0\,.
\ea
Again we emphasise that these properties will have to apply outside the original domain~\eqref{eq:DomainsStatic}.

\subsubsection{Example: Singling out generalised Fan-Wang regular black holes}\label{SecSubSub:FanWang}

Before we discuss a generalised example of regular black holes, let us first check that~\eqref{eq:Psidot} recovers the previous result for the Hayward regular black hole discussed in Sec.~\ref{SecSub:Hayward}. By comparing~\eqref{eq:PsiHayward} with~\eqref{eq:PsiHRho}, or alternatively by solving the metric function for the mass parameter $m$ and invoking~\eqref{eq:HRho} with~\eqref{eq:PsiHRho}, we obtain the generating function and its inverse
\ba
H\qty(\varrho) \,\,=\,\, \frac{\varrho}{1+ \ell^2 \varrho} \quad  \Leftrightarrow \quad   h(\psi) \,\,=\,\, \frac{\psi}{1-\ell^2 \psi} \,.
\ea
It is straightforward to verify that these satisfy all the assumptions detailed previously and associated limits on the  domains~\eqref{eq:DomainsStatic}. Moreover, $h(\psi)$ can be defined on the disjoint domain $\psi \in (\psi_\text{max},\infty)$, in which case it maps onto $\varrho \in \qty(-\infty,-\frac{1}{\ell^2})$. We do not consider the scenario where $\psi_+(v)$ becomes negative, as this is associated with a negative energy density of the shell and a discontinuity in the evolution at a finite time, cf.~Fig.~\ref{Fig:HaywardOriSecond2}.
For the above functions, we compute 
\ba\label{eq:hDhHDHHayward}
\frac{H\qty(\varrho)}{H'\qty(\varrho)} &=& \varrho \qty(1+\ell^2 \varrho)\,, \quad \quad  \frac{h(\psi)}{h'(\psi)} \,\,=\,\, \psi \qty(1- \ell^2 \psi) \,.
\ea
The latter ratio stays finite as $\psi$ crosses $\psi_{\text{max}}$ from below, so that the evolution equation~\eqref{eq:Psidot} can be applied in this regime even though $h(\psi)$ itself diverges.
Inserting the former expression into~\eqref{eq:Psidot} and focusing on the late-time regime, we neglect the linear term above within the considered approximation, because it makes a subleading contribution.
This leads to the differential equation
\ba
\dot{\psi}_+(v) & \simeq &  - \kappa_- \psi_+(v) + \frac{3 \ell^2 \lambda}{r_-v^{\gamma+1}} \psi_+(v)^2\,,
\ea
which is equivalent to~\eqref{eq:MdotHayward} taking into account the definition $\lambda = \frac{r_-^4}{6 \ell^2} \tilde{\lambda}$ and~\eqref{eq:PsiDef} with $R(v) \simeq r_-$. The general solution is given by
\ba
\psi_+(v) & \simeq &
 \frac{e^{-\kappa_- v}}{\frac{r_-^3}{2}c + \frac{r_-^3}{2}\tilde{\lambda} \kappa_-^\gamma\Gamma[-\gamma,\kappa_- v]} \,\, \simeq \,\, \frac{2}{r_-^3}\qty[\frac{ \kappa_-}{\tilde{\lambda}}v^{\gamma+1} + \mathcal{O}\qty(v^\gamma)]\,,
\ea
and we recover the previous result~\eqref{eq:MHaywardSol} of a late-time polynomial scaling following a transient period of mass inflation. In particular, we see that  a necessary consequence is that $\varrho_+(v)$ approaches a  quasi-fixed point in the limit $v\to \infty$ given by
\ba
\varrho_{+*} &=&- \frac{1}{\ell^2}  \quad \Leftrightarrow  \quad m_{+*} \,\,=\,\, - \frac{r_-^3}{2 \ell^2}\,.
\ea
This value is indeed singled out by the condition~\eqref{eq:RhoFPConditionH}. One may also verify that the universal late-time scaling~\eqref{eq:UniversalScaling} consistently recovers the previous result for the late-time expansion of $m_+(v)$ in~\eqref{eq:mExpansionHayward}.
As discussed in Sec.~\ref{SecSub:Hayward} and above, that $m_{+}(v)$ or equivalently $\varrho_+(v)$ becomes negative and approaches this fixed point is necessary to guarantee a divergent Misner-Sharp mass density for $v \to \infty$.
\\

We will now discuss a selected class of  limiting-curvature regular black holes for which this late-time power-law scaling can be made arbitrarily weak within the considered approximation.  One possibility to achieve this is if the theory-dependent quantity in the second square brackets of~\eqref{eq:Psidot2} for large $\psi_+(v)$ contributes by a correction term $\sim \psi_+(v)^{n+1}$ for $n \gg 1$, assuming again that an inverse $h_n\qty(\psi_+(v))$ can be defined for $\psi \in (\psi_{\text{max}},\infty)$. The previous Hayward case is recovered for $n=1$. On these grounds, let us consider a one-parameter family of generating functions $h_n(\psi)$ satisfying
\ba\label{eq:hPsiWant}
\frac{h_n(\psi)}{h'_n\qty(\psi)} &=& \psi \qty(1- \ell^{2 n} \psi^{n})\,.
\ea
Inserting this expression in~\eqref{eq:Psidot}, we again neglect the linear term above, because it makes a subleading contribution. This leads to the Bernoulli differential equation~\footnote{Such type of differential equations can be transformed to linear differential equations by substituting $\Psi_{n+}(v)\vcentcolon =1/ \psi_{n+}^{n}(v)$. For the solution to not encounter a singularity at a finite $v$, assuming $\psi_{n+}(v) >0$, the fact that $\lambda<0 $ is sufficient.}
\ba
\dot{\psi}_{n+}(v) & \simeq &  - \kappa_{n-} \psi_{n+}(v) + \frac{3 \ell^{2n} \lambda_n}{r_{n-}v^{\gamma+1}} \psi_{n+}(v)^{n+1}\,,
\ea
where $r_{n-}$ and $\kappa_{n-}$ are the inner-horizon radius and surface gravity, defined in~\eqref{eq:rMinus} and~\eqref{eq:kappaMinus}, of the yet-to-be-determined background metrics, and similarly $\lambda_n$ is computed according to~\eqref{eq:Lambda} and more specifically~\eqref{eq:LambdaH}. This also assumes that the background metrics are indeed capable of describing a double-horizon non-extremal regular black hole, which we will verify below.
Introducing the parameter $\tilde{\lambda}_n = \frac{6 \ell^{2n}}{r_{n-}^4 } \lambda_n < 0$ with length dimension $\gamma  - 3 + 2n$,
the general solution is
\ba\label{eq:PsinSol}
\psi_{n+}(v) & = & \qty[\frac{e^{-n \kappa_{n-} v}}{\frac{r_{n-}^3}{2}c + \frac{r_{n-}^3}{2} \tilde{\lambda}_n \kappa_{n-}^\gamma n^{\gamma+1}\Gamma\qty[-\gamma,n \kappa_{n-} v]}]^{\frac{1}{n}}
\,\,\simeq \,\, \qty[\frac{2}{r_{n-}^3}\qty[\frac{\kappa_{n-}}{\tilde{\lambda}_n} v^{\gamma+1} + \mathcal{O}\qty(v^{\gamma})]]^{\frac{1}{n}}\,,
\ea
where $c$ is an integration constant which, importantly, does not enter the asymptotic late-time expansion. Notice, moreover, that the leading term in square brackets is positive provided $\tilde{\lambda}_n < 0$, which holds by our assumptions on $\lambda_n$ according to~\eqref{eq:LambdaH}, and $\kappa_{n-}<0$, i.e., we are describing an inner horizon with negative surface gravity, and thus the fractional root is taken of a positive quantity.
Therefore, within the approximation considered here, for large but finite $n$, the late-time power-law scaling of the Misner-Sharp mass density of these yet-to-be-determined regular black holes can be made extremely weak. We will confirm this below by solving the exact system of equations~\eqref{eq:Mrdot1} and \eqref{eq:Mrdot2} numerically for two qualitatively different cases, extending the $n=1$ Hayward one discussed in Sec.~\ref{SecSub:Hayward}.\\

\begin{figure}[t]
	\centering
	\includegraphics[width=0.65\textwidth]{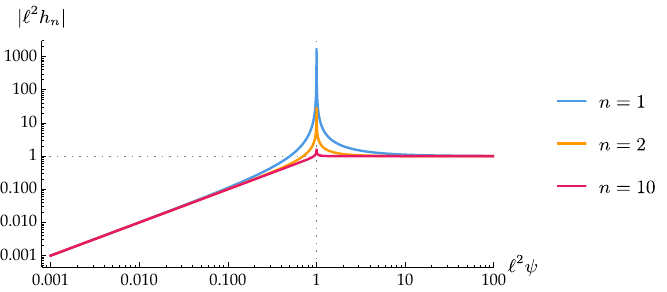} 
	\caption{\label{Fig:hnHn} Absolute value of the generating functions $h_n(\psi)$ in~\eqref{eq:hn} as functions of $\psi \in (0,\infty)$ for different $n$. The pole at $\psi_{\text{max}} = \frac{1}{\ell^2}$ is associated with the limiting value of the Misner-Sharp mass density of the static background spacetime. Transient mass inflation generally triggers an exponential growth of $\psi$ towards and beyond $\psi_{\text{max}}$, while the eternal late-time power-law divergence is reflected in a quasi-fixed point $ h_n(\psi) \to (-1)^{-\frac{1}{n}}\frac{1}{\ell^2}$.}
\end{figure}

To obtain the background metrics, we need the actual solution for the generating functions $h_n(\psi)$ satisfying~\eqref{eq:hPsiWant}. This first-order differential equation determines $h_n(\psi)$ up to a multiplicative factor, which on the domain $\psi \in (0,\psi_{\text{max}})$ we choose so that $h_n'(0)=1$ and in particular $h_n(\psi)$ is real and positive,
 whereby $\psi_{\text{max}} = \frac{1}{\ell^2}$ can be inferred by applying the condition~\eqref{eq:hPsihPrimemax} to~\eqref{eq:hPsiWant}, or alternatively directly from the solution below.
Thus, we obtain a one-parameter family of generating functions $H_n(\varrho)$ with inverses $h_n(\psi)$ on the domains~\eqref{eq:DomainsStatic} given by
\ba\label{eq:hn}
H_n\qty(\varrho) &=&  \frac{\varrho}{\qty[1 + \ell^{2n} \varrho^n]^{\frac{1}{n}}} \quad  \Leftrightarrow  \quad  h_n(\psi) \,\,=\,\, \frac{\psi}{\qty[1-l^{2n}\psi^n]^{\frac{1}{n}}} \,, \quad 
\ea
 where the defining relation~\eqref{eq:HRho} has been invoked. Fig.~\ref{Fig:hnHn} shows the modulus of $h_n(\psi)$ for different $n$ on the static domain $\qty(0,\psi_{\text{max}})$, as well as extended to the domain $\psi \in \qty(\psi_{\text{max}},\infty)$ where its definition must take into account an additional factor of $(-1)^{-\frac{1}{n}}$. Throughout  we adopt the principal branch for the fractional power in $h_n(\psi)$.  The Misner-Sharp density of the background spacetimes obtained by imposing~\eqref{eq:PsiHRho} is given by
 \ba\label{eq:Psin}
  \psi_n(m,r) &=&  \frac{2m}{\qty[r^{3n} + \qty(2m)^n \ell^{2n}]^{\frac{1}{n}}}\,,
 \ea
 where  $m>0$ is the asymptotic mass, and is a strictly monotonically decreasing function for $r\in(0,\infty)$. The
 corresponding spacetimes,
\ba\label{eq:fn}
f_n(m,r) &=&  1 - \frac{2m r^2}{\qty[r^{3n} + \qty(2m)^n \ell^{2n}]^{\frac{1}{n}}}\,,
\ea
are generalised Fan-Wang metrics~\cite{Fan:2016hvf,Tsuda:2026xjc}  describing a non-extremal double-horizon regular black hole provided $m > m_{\text{crit}} 
	=  \frac{1}{2}\qty[\frac{27}{4}]^{\frac{1}{2n}}  \ell  > 0$.
	This follows from the fact that $f_n(r)\to 1$ for $r\to \infty$ and $r\to 0$, as well as the fact that $f_n'(r)$ admits a single root for   $r\in(0,\infty)$ and $m>0$.\\

\begin{figure}[h!]
	\centering
	\includegraphics[width=0.73\textwidth]{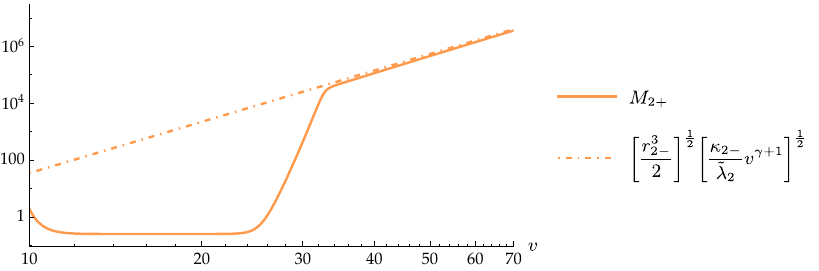}
	\caption{\label{Fig:FW2Ori} Misner-Sharp mass $M_{2+}(v)$ obtained by numerically integrating~\eqref{eq:Mrdot1} and~\eqref{eq:Mrdot2} for the Fan-Wang metric~\eqref{eq:fn} labeled by $n=2$ with free parameters set to $m_0 = 10$, $\ell = \frac{1}{2}$, $\beta = 1$, $\gamma=11$ and $v_0 =10$, and initial values $R(v_0)=1$ and $m_{2+}(v_0)=10+1$, in appropriate units of an external reference length scale, as well as analytically determined polynomial late-time scaling~\eqref{eq:PsinSol} following the transient mass-inflation period.}
\end{figure}

\begin{figure}[h!]
	\centering
	\includegraphics[width=0.65\textwidth]{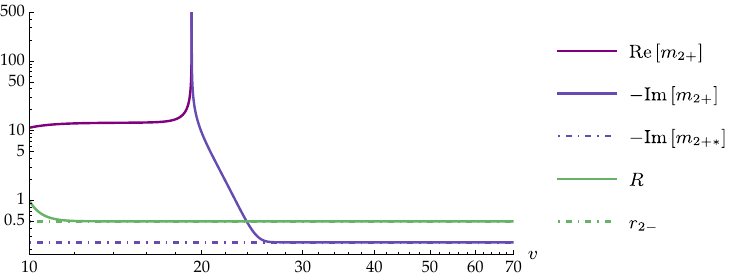}
	\caption{\label{Fig:FW2Ori2} Radial position of the shell $R(v)$ and mass parameter $m_{2+}(v)$ 	for the configuration corresponding to the choice of free parameters and initial values specified in Fig.~\ref{Fig:FW2Ori}. The dashed lines highlight the finite constant values $R(v)\simeq r_{2-}$ and $m_{2+}(v) \simeq m_{2+*}$ given by~\eqref{eq:mStarFW}, which are approached asymptotically in the limit $v\to \infty$.}
\end{figure}

\begin{figure}[h!]
	\centering
	\includegraphics[width=0.73\textwidth]{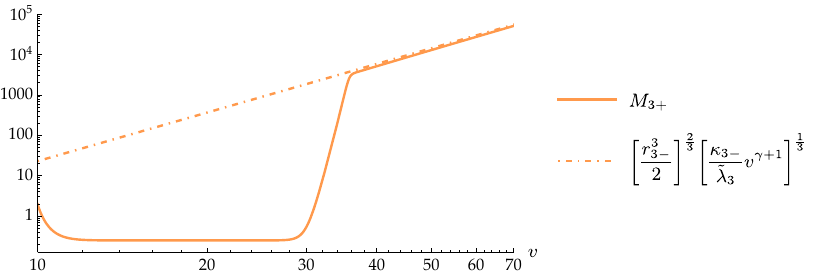}
	\caption{\label{Fig:FW3Ori} Misner-Sharp mass $M_{3+}(v)$ obtained by numerically integrating~\eqref{eq:Mrdot1} and~\eqref{eq:Mrdot2} for the Fan-Wang metric~\eqref{eq:fn} labeled by $n=3$ with free parameters set to $m_0 = 10$, $\ell = \frac{1}{2}$, $\beta = 1$, $\gamma=11$ and $v_0 =10$, and initial values $R(v_0)=1$ and $m_{3+}(v_0)=10+1$, in appropriate units of an external reference length scale, as well as analytically determined polynomial late-time scaling~\eqref{eq:PsinSol} following the transient mass-inflation period.}
\end{figure}

\begin{figure}[h!]
	\centering
	\includegraphics[width=0.65\textwidth]{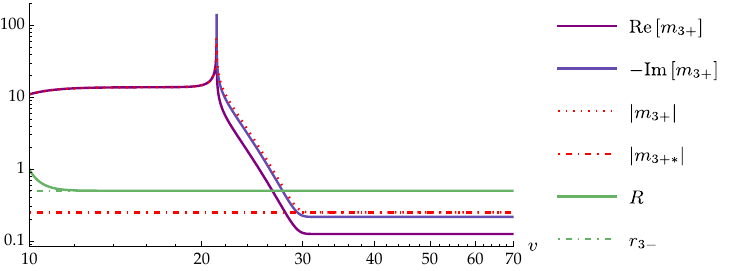}
	\caption{\label{Fig:FW3Ori2} Radial position of the shell $R(v)$ and mass parameter $m_{3+}(v)$ 	for the configuration corresponding to the choice of free parameters and initial values specified in Fig.~\ref{Fig:FW3Ori}. The dashed lines highlight the finite constant values $R(v)\simeq r_{3-}$ and $m_{3+}(v) \simeq m_{3+*}$ given by~\eqref{eq:mStarFW}, which are approached asymptotically in the limit $v\to \infty$.}
\end{figure}

\begin{figure}[h!]
	\centering
	\includegraphics[width=0.62\textwidth]{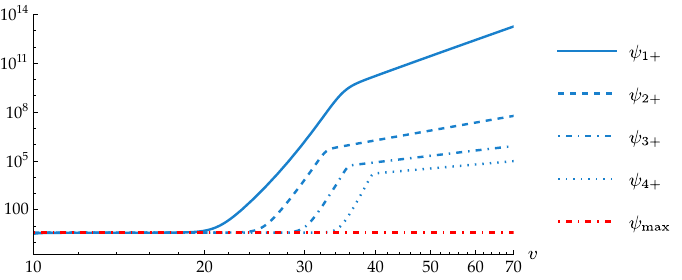}
	\caption{\label{Fig:FWnOri2} Misner-Sharp mass densities $\psi_{n+}(v)$ for the configurations corresponding to the choice of free parameters and initial values identical as in Fig.~\ref{Fig:HaywardOri}, for different $n$. The mass-inflation regime is squinched for large $n$ and the late-time power-law scaling is weakened.}
\end{figure}

	At the same time, we see that the asymptotic late-time power-law divergence of $\psi_{n+}(v)$ of the perturbed spacetime pushes $\varrho_{n+}(v)$ to a (set of possible) quasi-fixed point(s) given  by
	\ba\label{eq:mStarFW}
	\varrho_{n+*} &=&  \qty(-1)^{-\frac{1}{n}}\frac{1}{\ell^2} \quad \Leftrightarrow  \quad m_{n+*} \,\,=\,\, \qty(-1)^{-\frac{1}{n}} \frac{r_{n-}^3}{2 \ell^2}\,.\quad 
	\ea
In particular this unphysical parameter, or equivalently $m_{n+}(v)$, necessarily becomes negative for $n=1$, imaginary for $n=2$, and complex with non-vanishing real and imaginary parts for finite $n>2$ as $v\to \infty$. This is required in order for the physical  Misner-Sharp mass density $\psi_{n+}(v)$  to become large-positive divergent. Again, we see that this divergence is singled out by the corresponding curvature singularity of the background spacetimes~\eqref{eq:fn} taylored to $r=r_{n-}$, with associated values of $m_{n+*}$ which can be determined by the condition~\eqref{eq:RhoFPConditionH}, i.e., by the roots of 
\ba
\frac{H_n\qty(\varrho)}{\varrho H_n'(\varrho)} &=& 1+\ell^{2n} \varrho^n\,.
\ea
To confirm that this late-time power law regime is  reached by suitably chosen initial conditions, we solve the exact equations~\eqref{eq:Mrdot1} and~\eqref{eq:Mrdot2} numerically for two qualitatively different cases $n=2$ and $n=3$, whereby the quantity defined in~\eqref{eq:B} is given by
\ba
B_n\qty(M_{n+}(v),R(v)) &=& 3 \cdot 2^n \ell^{2n}  \frac{M_{n+}(v)^{n+1}}{R(v)^{3 n+1}}\,.
\ea
Throughout we use the same choice of free parameters and initial conditions as for the Hayward spacetime corresponding to $n=1$ in Fig.~\ref{Fig:HaywardOri} and Fig.~\ref{Fig:HaywardOri2}.
The result for $M_{2+}(v)$ and $M_{3+}(v)$ is shown in Fig.~\ref{Fig:FW2Ori} and Fig.~\ref{Fig:FW3Ori}, where we also compare these with the universal late-time scaling extracted from~\eqref{eq:PsinSol} using the definition~\eqref{eq:PsiDef}. For fixed initial conditions and increasing $n$, the transition between exponential and polynomial growth occurs at lower orders of magnitude of $\psi_{n+}(v)$, indicating that for large $n$ mass inflation can be effectively tamed. In Fig.~\ref{Fig:FW2Ori2} and Fig.~\ref{Fig:FW3Ori2} we show the corresponding behavior of the unphysical mass variables $m_{2+}(v)$ and $m_{3+}(v)$.~\footnote{For these plots we have solved $M_{n+}\qty(m_{n+}(v),R(v))$ for $m_{n+}\qty(M_{n+}(v),R(v))$ and chosen the principal solution branch.}  This is to confirm that, while for $n=1$ this variable enters a disjoint domain of negative real values, cf.~Fig.~\ref{Fig:HaywardOri}, for $n=2$ it switches to purely imaginary, while for $n=3$ it becomes complex with non-vanishing real and imarginary parts. In Fig.~\ref{Fig:FWnOri2} we moreover show the Misner-Sharp mass densities $\psi_{n+}(v)$ for different $n$ and identical initial conditions to illustrate the squinching of mass inflation for increasing $n$. Different from the Misner-Sharp masses $M_{n+}(v)$, the densities $\psi_{n+}(v)$ are strictly monotonically increasing at all times. \\


\newpage For completeness, let us emphasise that reality of the interior metrics for $\psi_{n+}(v) \in ( \psi_{\text{max}},\infty )$ can be observed explicitly by choosing the alternative parametrisation
\ba
\qty(2 m_{n+}(v))^n \ell^{2n} &= \vcentcolon & -s_{n+}^{3n}(v)\,,
\ea
where the variables $s_{n+}(v)$ are real. Herewith the interior Misner-Sharp masses $M_{n+}(v)$ in this regime are given by
\ba
M_{n+}(v) &=& \frac{s_{n+}^3(v) R(v)^3}{2 \ell^2 \qty[s_{n+}^{3n}(v) - R(v)^{3n}]^{\frac{1}{n}}} \,.
\ea
These expressions are manifestly real for $R(v) < s_{n+}(v)$, and the singularity approached in the limit $v\to \infty$ is described by $R(v) \to r_{n-}$ and $s_{n+}(v) \to  r_{n-}$. In particular, this gives the result~\eqref{eq:mStarFW} raised to the power $n$ and avoids the discussion of complex $m_{n+}(v)$.  We have nevertheless decided to discuss the feature of complex $m_{n+}(v)$ for the following reason. The constant parameters $m$ in the static line elements~\eqref{eq:fn} from the perspective of vacuum solutions to quasi-topological gravities arise as integration constants in the integrated equation of motion~\eqref{eq:hmAlgebraic} which determines the background metrics~\cite{Bueno:2024dgm,Bueno:2025qjk,Bueno:2025zaj,Borissova:2026krh,Borissova:2026wmn}. We expect an extension of our discussion in the context of an analogue Ori model based on the junction conditions for quasi-topological gravities~\cite{Bueno:2026wjt} to answer important questions about the higher-curvature resummation series determining the function $h(\psi)$, its radius of convergence and dynamical evolution under perturbations.

\subsubsection{Other example of a generating function}\label{SecSubSub:Other}

As a different type of example, we consider the following generating function,
\ba
h(\psi) &=& \frac{\psi}{\qty[1-\ell^2 \psi]^2} \,.
\ea
This function satisfies all the previous assumptions on the domain $\psi \in (0,\psi_{\text{max}})$. 
If we compute the ratio between this function and its derivative, and herewith check the condition~\eqref{eq:Massinflationeternal} by expanding the result for  large $\psi$, we obtain
\ba\label{eq:hRatioExample2}
\frac{h\qty(\psi)}{h'\qty(\psi)} &=& \frac{\psi - \ell^2 \psi^2}{1 +\ell^2 \psi} \,\,=\,\, -\psi + \mathcal{O}\qty(\psi^0)\,.
\ea
This is linear to leading order, and 
so we expect that for the corresponding regular black hole  mass inflation will be eternal. \\

\begin{figure}[h!]
 	\centering
 	\includegraphics[width=0.62\textwidth]{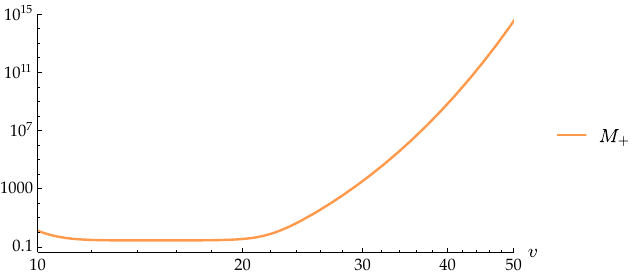}
 	\caption{\label{Fig:OtherOri} Misner-Sharp mass $M_{+}(v)$ obtained by numerically integrating~\eqref{eq:Mrdot1} and~\eqref{eq:Mrdot2} for the minus metric~\eqref{eq:f2pmOther} with free parameters set to $m_0 = 10$, $\ell = \frac{1}{2}$, $\beta = 1$, $\gamma=11$ and $v_0 =10$, and initial values $R(v_0)=1$ and $m_{+}(v_0)=10+1$, in appropriate units of an external reference length scale.}
 \end{figure}

\begin{figure}[h!]
 	\centering
 	\includegraphics[width=0.65\textwidth]{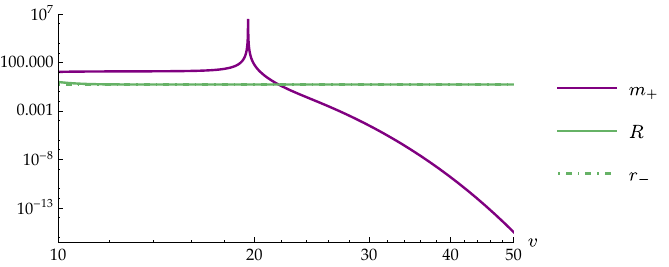}
 	\caption{\label{Fig:OtherOri2} Radial position of the shell $R(v)$ and mass parameter $m_{+}(v)$ for the configuration corresponding to the choice of free parameters and initial values specified in Fig.~\ref{Fig:OtherOri}.}
 \end{figure}

We determine the background metric by invoking~\eqref{eq:hmAlgebraic} and~\eqref{eq:HRho}. This requires finding the roots of a quadratic polynomial in $\psi$, which has two solution branches
\ba
H^\pm(\varrho) &=& \frac{1 + 2 \ell^2 \varrho \pm \sqrt{1+ 4 \ell^2 \varrho}}{2 \ell^4 \varrho}\,,
\ea
with corresponding metrics explicitly
\ba\label{eq:f2pmOther}
f^\pm(m,r) 
&=& 1 -\frac{r^2}{\ell^2} - \frac{r^5}{4 m \ell^4} \qty[1 \pm \sqrt{1+ \frac{8 m \ell^2}{r^3}}]\,.
\ea
Here $f^-(m,r)$ is the asymptotically flat regular black hole solution branch  which provides the physical background, and the starting point for the numerical integration
of~\eqref{eq:Mrdot1} and~\eqref{eq:Mrdot2}.
Hereby  the quantity defined in~\eqref{eq:B} is given by
	\ba
	B\qty(M_{+}(v),R(v)) &=&  \frac{12 \ell^2}{R(v)}\frac{ M_+(v)^2}{R(v)^3 + 2 \ell^2 M_+(v)}\,,
	\ea
where we have used~\eqref{eq:BPsi} together with~\eqref{eq:hRatioExample2}. The exact equation for $M_+(v)$ remains regular as $\psi_+(v)$ crosses $\psi_{\text{max}}$ from below. The pole of $B\qty(M_{+}(v),R(v)) $ at negative $M_+(v) = - \frac{R(v)^3}{2\ell^2}$ is not encountered for our choice of free parameters and initial conditions, which throughout we 
 take to be identical as for the Fan-Wang class.
 The result for $M_{+}(v)$ is shown in Fig.~\ref{Fig:OtherOri} and indeed displays an eternal exponential divergence. At the same time, the unphysical mass parameter $m_{+}(v)$ encounters the expected divergence at a finite time as the solution crosses the limiting value of Misner-Sharp mass density of the background. Thereafter it continues to range in the domain $m_+(v) \in (0,\infty)$ and asymptotically goes to zero from above, $m_{+}(v) \searrow 0$ for $v\to \infty$, cf.~Fig.~\ref{Fig:OtherOri2}. Importantly, a physical divergence $M_+(v) \to \infty$ at this value of $m_{+}(v)$ could not have happened on the original solution branch $f^-_{+}\qty(m_+(v),R(v))$. Indeed, expanding~\eqref{eq:f2pmOther} for small $m$ gives
 \ba
 f^-(m,r) &=& 1 + \mathcal{O}(m)\,,\\
 f^+(m,r) &=& - \frac{r^5}{2 \ell^4 m} +  \mathcal{O}\qty(m^0)\,.
 \ea
The second solution branch is the one that has a physical singularity for finite $r$ and $m\to 0$. This is the singularity that is opened up by mass inflation.
 In other words, mass inflation effectively forces a dynamical non-flat extension of the static spacetime. This is a phenomenon distinct from the previously discussed examples as well as other examples discussed in the literature~\cite{Carballo-Rubio:2021bpr,Bonanno:2025bgc} which deserves further investigation beyond the scope of this work.
 
\newpage
\section{Conclusion}\label{Sec:Discussion}

{\it If} the perturbation of static non-extremal  de Sitter core regular black holes with asymptotic mass $m > 0$  and an $m$-independent upper bound on the curvature results in a divergence of the physical Misner-Sharp mass density,
then this divergence within models based on Vaidya generalisations $m\mapsto m(v)$ of the associated static line elements can be understood largely geometrically. For instance, it may require that a given unphysical mass parameter $m(v)$ of the model evolves outside the positive-real domain, or that the static metric is dynamically continued to a spacetime which violates the assumption of limiting curvature even for $m(v) > 0$. {\it That} this unbounded growth of physical  quantities occurs is the essence of mass inflation discovered originally for a singular charged black hole in general relativity~\cite{Poisson:1989zz,Poisson:1990eh}, and discussed for regular black holes particularly in~\cite{Brown:2011tv,Carballo-Rubio:2018pmi,Bonanno:2020fgp,Carballo-Rubio:2021bpr,Bonanno:2022rvo,Carballo-Rubio:2022pzu,Bonanno:2023qhp,Bonanno:2025bgc}.  

 We have addressed this conjecture dynamically
  for a  class of limiting-curvature regular black holes inspired by quasi-topological gravity within the generalised Ori model~\cite{Ori:1991zz}, in which these spacetimes are realised as effective solutions of general relativity by making use of the junction conditions for a null thin shell modeling the outgoing perturbation. Our discussion supports the viewpoint that an eternal late-time divergence of the physical Misner-Sharp mass density 
 is associated with the curvature singularity of an appropriate dynamical real extension of the static spacetime. 
 
 A key question is whether this, in principle, purely kinematic statement extends to quasi-topological gravitational theories giving rise to limiting-curvature regular black holes as vacuum solutions generically~\cite{Bueno:2024dgm,Frolov:2024hhe,Bueno:2025tli,Bueno:2024zsx,Bueno:2025zaj,Borissova:2026wmn,Borissova:2026krh}.
Mass inflation based on double null thin-shell models in higher-dimensional quasi-topological gravities has been addressed in~\cite{Frolov:2026rcm,DiFilippo:2026jpv,Bueno:2026wjt}. With the tools developed in~\cite{Bueno:2026wjt}, it should be possible to address the same questions asked here based on an analogue Ori model using the junction conditions for a null thin shell in quasi-topological gravity. In particular, it will be interesting to understand whether the infinite resummation series resulting in functions $h(\psi)$ associated with de Sitter core regular black holes~\cite{Bueno:2024dgm} is indeed pushed out of its radius of convergence dynamically, whether this characteristic function can take values outside the positive-real domain 
while the metric stays real but ceases to have limiting curvature, and what this would imply for the regime of validity of the underlying higher-curvature gravitational theory --- or alternatively, whether instead the fact that the vacuum theory prescribes strictly real and positive $h(\psi) $ could imply, within an analogue Ori model for quasi-topological gravity, that mass inflation is capped at the limiting Misner-Sharp mass density of the background.

Finally, let us also emphasise the fact that the Ori null thin-shell model may fail to capture the full nature of the inner-horizon instability and leading-order divergences of curvature invariants in the presence of continuous crossflow perturbations~\cite{Poisson:1989zz,Poisson:1990eh,Brady:1992cz,BorissovaToAppear}. 

\begin{acknowledgments}
	
I wish to thank Ra\'ul Carballo-Rubio and Antonio Panassiti for insightful discussions. I am also grateful to Alfio Bonanno for useful comments on a previous version of the manuscript.
 This work is supported by STFC Consolidated Grant ST/X000575/1 and STFC Consolidated Award UKRI4230.

\end{acknowledgments}

\enlargethispage{20pt}

\bibliographystyle{jhep}
\bibliography{references}

\end{document}